%% file: MOO_global_tracking.tex
\documentclass[fleqn,10pt]{wlscirep}
\usepackage{amsmath,amssymb}

\usepackage{graphicx}
\usepackage{subcaption}
\usepackage[utf8]{inputenc}
\usepackage[export]{adjustbox}
\usepackage{wrapfig}
\usepackage{float}
\usepackage{xcolor,colortbl}
\usepackage{subfiles}

\newcommand{\beginsupplement}{%
        \setcounter{table}{0}
        \renewcommand{\thetable}{S\arabic{table}}%
        \setcounter{figure}{0}
        \renewcommand{\thefigure}{S\arabic{figure}}%
     }

\title{Optimization and Validation of Diffusion MRI-based Fiber Tracking with Neural Tracer Data as a Reference}

\author[1*]{Carlos Enrique Gutierrez}
\author[2]{Henrik Skibbe}
\author[3]{Ken Nakae}
\author[1]{Hiromichi Tsukada}
\author[1]{Jean Lienard}
\author[4]{Akiya Watakabe}
\author[5,9,10]{Junichi Hata}
\author[6]{Marco Reisert}
\author[7]{Alexander Woodward}
\author[5,10]{Hideyuki Okano}
\author[4]{Tetsuo Yamamori}
\author[8,11,12]{Yoko Yamaguchi}
\author[3]{Shin Ishii}
\author[1]{Kenji Doya}

\affil[1]{Neural Computation Unit, Okinawa Institute of Science and Technology Graduate University, Okinawa, Japan}
\affil[2]{Brain Image Analysis Unit, Riken Center for Brain Science, Wako, Japan}
\affil[3]{Integrated Systems Biology Laboratory, Department of Systems Science, Graduate School of Informatics, Kyoto University, Kyoto, Japan}
\affil[4]{Laboratory for Molecular Analysis of Higher Brain Function, RIKEN Brain Science Institute, Wako, Japan}
\affil[5]{Laboratory for Marmoset Neural Architecture, RIKEN Brain Science Institute, Wako, Japan}
\affil[6]{Department of Medical Physics, Medical Center, Freiburg University, Germany}
\affil[7]{Connectome Analysis Unit, Riken Center for Brain Science, Wako, Japan}
\affil[8]{Applied Electronics Laboratory, Kanazawa Institute of Technology, Japan}
\affil[9]{Division of Regenerative Medicine, The Jikei University School of Medicine, Tokyo, Japan}
\affil[10]{Department of Physiology, Keio University School of Medicine, Tokyo, Japan}
\affil[11]{Graduate School of Information Science and Technology, The University of Tokyo, Tokyo, Japan}
\affil[12]{Laboratory for Cognitive Brain Mapping, RIKEN Center for Brain Science, Wako, Japan}

\affil[*]{carlos.gutierrez@oist.jp}

\begin{abstract}
Diffusion-weighted magnetic resonance imaging (dMRI) allows non-invasive  investigation of whole-brain connectivity, which can potentially help to reveal the brain's global network architecture and abnormalities involved in neurological and mental disorders. However, the reliability of connection inferences from dMRI-based fiber tracking is still debated, due to low sensitivity, dominance of false positives, and inaccurate and incomplete reconstruction of long-range connections. Furthermore, parameters of tracking algorithms are typically tuned in a heuristic way, which leaves room for manipulation of an intended result.
Here we propose a data-driven framework to optimize and validate  parameters of dMRI-based fiber-tracking algorithms using neural tracer data as a reference. Japan's Brain/MINDS Project provides invaluable datasets containing both dMRI and neural tracer data from the same primates. A fundamental difference when comparing the dMRI-based tractography and neural tracer data is that the former cannot specify the direction of connectivity; therefore, evaluating the fitting of dMRI-based tractography becomes challenging. We considered four criteria for goodness of fiber tracking: distance-weighted coverage, true/false positive ratio, projection coincidence, and commissural passage, applied using a multi-objective optimization algorithm. We implemented a variant of  non-dominated sorting genetic algorithm II (NSGA-II) to optimize five major parameters of a global fiber-tracking algorithm over multiple brain samples in parallel.
Using optimized parameters compared to the default parameters, dMRI-based fiber tracking performance was significantly improved: average fiber length from 10mm to 17mm, voxel-wise coverage of axonal tracts from 0.9\% to 15\%, and the correlation of target areas from 40\% to 68\%, while minimizing false positives and impossible cross-hemisphere connections.
Parameters optimized for 10 tracer injection sites showed good generalization capability for other brain samples. These results demonstrate the importance of data-driven adjustment of fiber-tracking algorithms and support the validity of dMRI-based tractography, if appropriate adjustments are employed.
\end{abstract}

\begin{document}

\maketitle

\thispagestyle{empty}

\section*{Introduction}
Diffusion-weighted magnetic resonance imaging (dMRI) generates images based on the anisotropic diffusion of water molecules. Diffusion in the brain is constrained in a direction-dependent manner by obstacles such as nerve fibers and membranes. This leads to anisotropic diffusion patterns in dMRI images that can be used to estimate structural brain connectivity in a non-invasive way \cite{moseley1990diffusion, conturo1999tracking, mori1999three, basser2000vivo, tuch2002high}. 
dMRI-based tractography can trace whole-brain connectivity to more fully reveal network organization\cite{sporns2010networks, sporns2013structure, rubinov2010complex}, its relationship with functions\cite{passingham2002anatomical, schmahmann2007complex, chedotal2010wiring}, mental and neurological disorders \cite{bassett2009human, stam2014modern, xue2014diffusion, skudlarski2010brain}, and computational modeling\cite{izhikevich2008large}.
However, there are fundamental limitations, such as the lack of directionality of connections and the difficulty of estimating crossing fiber orientations in voxels of low spatial resolution\cite{fillard2011quantitative, schilling2017can}.
Other practical issues include:
{
\begin{enumerate}
    \item Low sensitivity, or low true positive (TP) rate \cite{calabrese2015diffusion, thomas2014anatomical, zalesky2016connectome}. 
    \item Low specificity, or high false positive (FP) rate \cite{thomas2014anatomical, drakesmith2015overcoming, maier2017challenge}.
    \item Difficulty to track long-distance connections \cite{reveley2015superficial, donahue2016using, sinke2018diffusion}.
\end{enumerate}
}
Unfortunately, all of these potentially contribute to erroneous reconstruction of connectomes.

Various efforts have been made to improve the accuracy of reconstructions. Global tractography\cite{reisert2011global, mangin2013toward, christiaens2015global} provides whole-brain connectivity that consistently explains dMRI data by optimizing a global objective function. 

Compared to conventional seed-based fiber tracking, it achieved better qualitative results on phantom data\cite{reisert2011global}.
However, both seed-based and global fiber tracking algorithms have a number of parameters that are difficult to determine because of unknown biophysical variables.

Japan's Brain/MINDS project (Brain Mapping by Integrated Neurotechnologies for Disease Studies)\cite{okano2016brain} intends to build a multi-scale marmoset brain map and mental disease models. 
The project has assembled a high-resolution marmoset brain atlas\cite{woodward2018brain} and is
conducting systematic anterograde tracer injections to analyse brain connectivity, while obtaining functional, structural, and diffusion MRI for most individuals.
All data are mapped to a common brain space.
This gives us a unique opportunity to verify the accuracy of dMRI-based fiber tracking using neuronal tracer data, reconstructed with the marmonet pipeline\cite{skibbe2019marmonet} as a reference.

Here we propose a  multi-objective optimization approach to evaluate the results of global fiber tracking from dMRI data with different parameter settings in reference to neuronal tracer data from multiple injection sites.
We consider four objective functions to target issues 1), 2) and 3): i) distance-weighted coverage, ii) the true/false positive ratio, iii) projection coincidence, and iv) commissural passage.

We take the five major parameters of the global tracking algorithm\cite{reisert2011global} and apply a multi-objective optimization algorithm based on non-dominated sorting genetic algorithm II (NSGA-II)\cite{deb2002fast} to find parameters that perform well for multiple brain samples.
We optimize the parameters using 10 brain samples and then test their capacity for generalization using 6 brain samples that were not used for optimization.

The developed code is compatible with HPC (high-performance computing) clusters to process multiple brain samples in parallel and the code is publicly available.

Our multi-objective optimization framework can be applied to other fiber-tracking algorithms or other objective functions, such as using micro-scale axonal orientation information from myelin stain data to improve reconstruction accuracy\cite{zhang2018optimization}. 

\section*{Results}

First, we consider criteria applied for evaluation of dMRI-based fiber tracking and the parallel multi-objective optimization method to find parameters that perform best for multiple samples with respect to our objective functions.
We then compare the results of fiber tracking with and without optimization and test how the optimized parameters generalize for test samples.

\subsection*{Criteria for evaluation}
Straightforward criteria for evaluation include sensitivity (true positive rate), specificity (true negative rate), and precision (true positive/all positive).
Fitting can be quantified for axon trajectories at the voxel level or for projection targets at the brain-region level.
An important issue in comparing dMRI-based tracking and anterograde neural tracer data is that the former does not reflect the projection direction.
dMRI-based fibers connected to a tracer injection site can include both incoming and outgoing axons to the site.
Thus, if we take anterograde tracing as a reference, it is natural to have additional "false positive" fibers.
Another issue in dMRI-based fiber tracking is the difficulty of tracking long connections, such as cross-hemisphere or sub-cortical connections.

\begin{figure}[h!]
\centering
\includegraphics[scale=0.295]{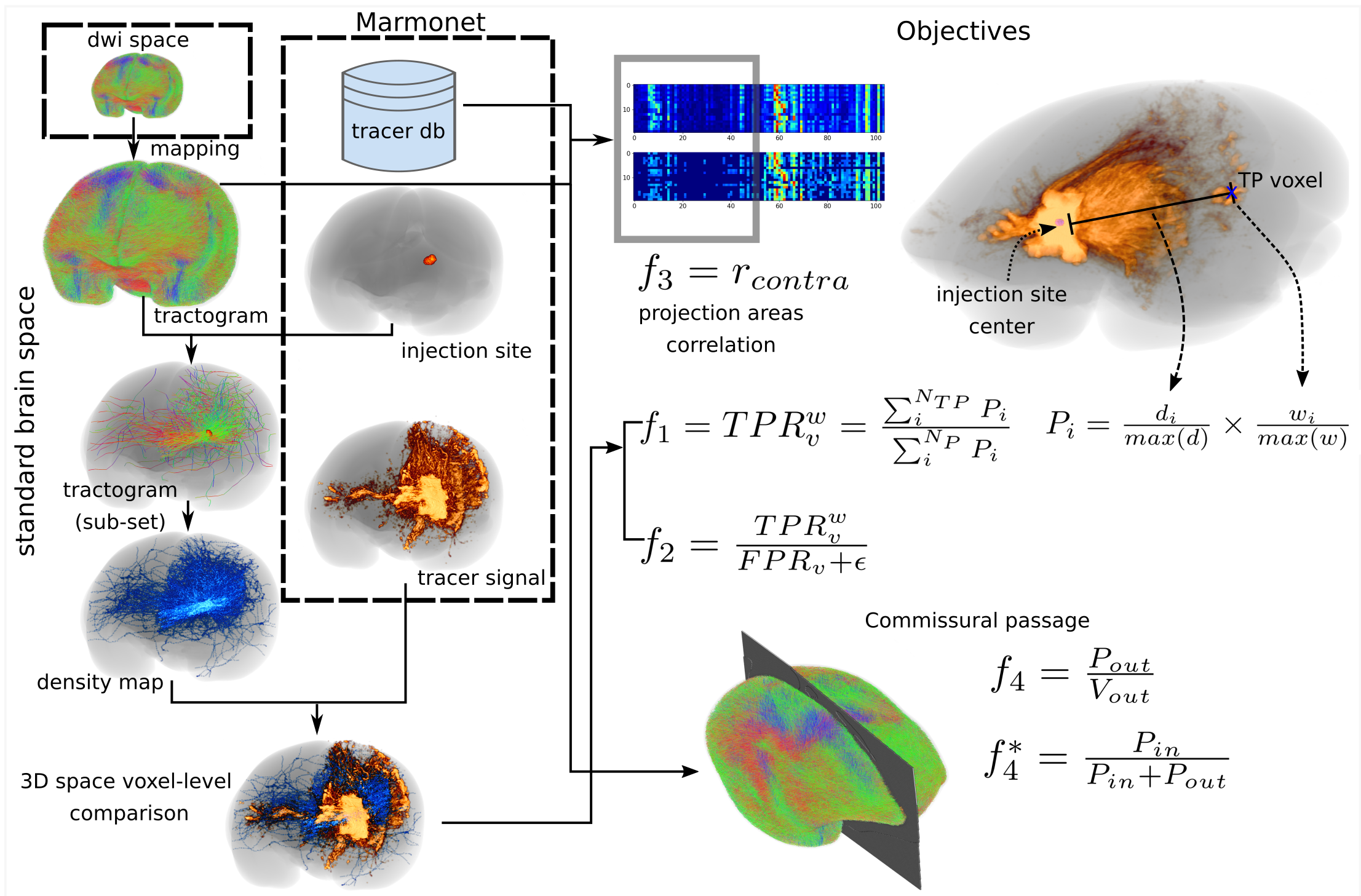}
\caption{\textbf{Criteria for evaluation.} Global fiber tracking results are mapped to the standard brain space and intersected spatially with the injection site, allowing extraction of a subset of fibers. The full tractogram is used to compute projection coincidence with the target hemisphere $f_3$ and the commissural passage $f_4$, while a subset of fibers is used for the distance-weighted coverage $f_1$ and true/false positive ratio $f_2$ objectives. True positive voxels are weighted by 2 factors extracted from neural tracer data, the distance to the injection site center $d_i$ and the voxel intensity $w_i$.}
\label{fig:criteria}
\end{figure}

Accordingly, we consider the following four objective functions (Fig.\ref{fig:criteria}): i) distance-weighted coverage, ii) the true/false positive ratio, iii) projection coincidence, and iv) commissural passage, as explained below.

\begin{description}
\item[i) Distance-weighted coverage] $f_1=TPR^w_v=\frac{\sum_i^{N_{TP}}P_i}{\sum_i^{N_P}{P_i}}$. 
Here, $P_i=\frac{d_i}{max(d)} \times \frac{w_i}{max(w)}$ is a positive voxel in the 3D tracer image reconstruction that is weighted by voxel fluorescent intensity $w_i$ and the distance $d_i$ from the voxel to the center of the injection region. 
This objective is maximized and uses $d_i$ and $w_i$ to promote long-range connections, with voxels strongly connected to the injection region. $N_{TP}$ is the total number of true positive voxels found in the comparison, and $N_P$ the total number of positive voxels in the tracer data.
\item[ii) True/false positive ratio]
$f_2=\frac{TPR^w_v}{FPR_v+\epsilon}$. Here, $FPR_v$ is the false positive rate at the voxel-level, and $\epsilon$ is the tolerance term calculated empirically and given by $\epsilon=0.006\times\frac{\mu_P}{\mu_N}$, with $\mu_N$ equal to the average number of true negative $TN$ voxels in individual whole-brain masks for the training data set, and $\mu_P$, similarly, the mean number of true positive $TP$ voxels. $\mu_N$ is a large number. $\epsilon$ provides the minimum acceptable value of $FPR_v$, considering for example, that tractography results would be adequate, even if up to 0.6$\%$ of the $TP$ are missed and counted as $FP$. Our optimization used $\epsilon=0.0013$. Maximization of this objective drives $TPR^w_v$ growth, while maintaining $FPR_v$ below a reasonable level, helps to constrain the dominance of $FP$\cite{maier2017challenge}. We observed cases in which small increments of $FPR_v$ resulted in the maximization of ii), thus, we added i) cost explicitly to adjust ii) in the right direction.
\item[iii) Projection coincidence]
$f_3=r_{contra}$, the Spearman rank correlation coefficient between neural tracer and dMRI tractography-based connectome matrices for the contralateral-hemisphere of the brain.
This objective function promotes accuracy of long cross-hemisphere projections.
Global tractography was run twice with the same parameters, and results were averaged and mapped to the tracer-based connectome matrix\cite{skibbe2019marmonet} of 20 sources $\times$ 104 targets parcellation. Both matrices were log-normalized.

\item[iv) Commissural passage]
$f_4=\frac{P_{out}}{V_{out}}$.
While direction-insensitive dMRI fiber tracking should yield many "false positives" in reference to anterograde neural tracers, some of the estimated paths can be impossible, such as those crossing hemispheres outside of commissural areas.

This criterion uses a binary mask in the midline, covering voxels outside anatomical commissures, such as the corpus callosum and the cerebellum.
$P_{out}$ is the number of voxels for fibers crossing the mid-line outside commissures, 
and
$V_{out}$ is the total number of positive voxels of the mask. 
This objective is targeted for minimization, and supports the non-dominance of $FP$. $f_4$ is additionally evaluated as $f_4^*=\frac{P_{in}}{P_{in}+P_{out}}$, where $P_{in}$ counts the voxels of fibers passing through the anatomical commissures. $f_4^*$ provides the proportion of true anatomical reconstructions at the commissures and the accuracy of optimization for interconnection of the two sides of the brain.

\end{description}

\subsection*{dMRI-based fiber tracking and its parameters}
Here we take the global fiber tracking algorithm\cite{reisert2011global} for our implementation, but the framework can be used with any other dMRI-based fiber tracking algorithms.
We explore the major parameters $\theta$ = [width $\sigma$, length $l$, weight $w$, chemPot $c$, connlike $L$]\cite{reisert2011global} (see Global tractography and parameters selection in methods). 

\subsection*{Parallel multi-objective optimization}
\begin{figure}[h]
\centering
\includegraphics[scale=0.32]{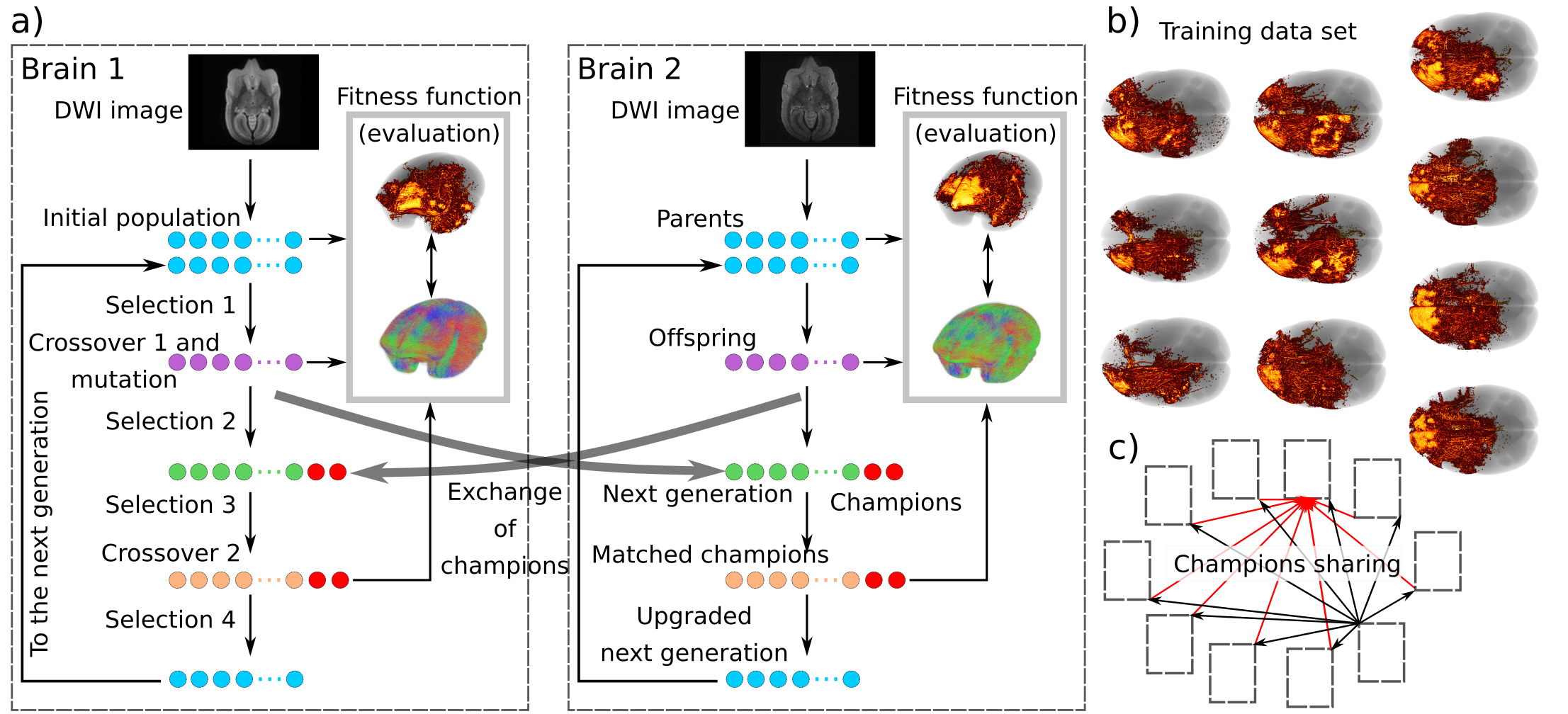}
\caption{\textbf{Multi-objective optimization (MOO) process.} a) From the initial population of parameters (parents, light blue dots), fitness values are obtained. Tournament selection (selection 1) creates offspring (purple dots). Crossover and mutation are performed on offspring and fitness values are calculated. From the combined set of parents and offspring, selection of the best elements (selection 2) creates the next generation (green dots). The best elements are shared among brain optimization processes (red dots). Most dominant elements (selection 3) are taken from the next generation and mixed with champions via a crossover operation. After obtaining objective values for matched elements and original champions, the next generation is upgraded by selecting the best elements from the joint set "next generation + original champions + matched champions" and sent as a parent for the next iteration. b) One MOO runs for each brain of the training data set, c) sharing the $i$-iteration champion to all MOO-processes (black arrows) and receiving its champion as well (red arrows).}
\label{fig:process_moo}
\end{figure}

The goal of multi-objective optimization (MOO) is to approximate the globally pareto-optimal set, also known as the pareto front, where no objective function values can be improved without degrading some other objective values. Solutions provided are non-dominated among them (dominance conditions of solution $x$ over solution $y$ are: $x$ is no worse than $y$ in any objective, and $x$ is better than $y$ in at least 1 objective) and evenly distributed\cite{deb2001multi, deb2002fast, coello2006twenty}. The following process is applied to each brain, as shown in Fig.\ref{fig:process_moo}. For each loop of the optimization process, objectives i), ii), iii) and iv) are integrated as a 4D-fitness value within NSGA-II\cite{deb2002fast}, and optimized until convergence. 

Parameters are initialized to their default values: $\mu_{\theta}=[0.1,0.3,0.133,0.2,0.5]$\cite{reisert2011global}, and the exploration is defined within heuristically determined lower $[0.01,0.24,0.01,0.05,0.5]$ and upper $[0.15,0.65,0.22,0.6,6.0]$ bounds. A population $M$ of size 8 where each element $M_i$, called an $individual$, is an array of length 5, corresponding to the parameters to optimize $\theta$, and is drawn from random uniform distributions with mean $\mu_{\theta}$ and standard deviation $\sigma_{\theta}=0.01$ except for $\sigma_{weight}=0.001$. Fitness values of the initial population $M$ (objectives) are calculated and the generational NSGA-II-based process begins.\\ 

Depending on fitness values, tournament, dominance-based selection between 2 individuals $M_i$ is performed. If the $f(M_i)$'s pair does not inter-dominate, selection is accomplished by evaluating the crowding distance\cite{deb2002fast}. With repetition, the tournament selects 8 offspring. We choose to invalidate the fitness of the offspring and perform crossover and mutation directly. Crossover picks individuals at even positions of the offspring array and pairs them with individuals in odd positions. Crossover uses simulated binary crossover \cite{agrawal1995simulated}, which is applied to each pair with probability $cxp=0.2$ of matching two individuals. Mutation is applied to all individuals among the offspring using a polynomial approach\cite{agrawal1995simulated}. 
Offspring fitness values are calculated. Then, from the combined set of parents and offspring, the next generation of 8 elements is selected based on fitness values and spread\cite{deb2002fast}.

In addition, the best individual is selected from the combined set as the local "champion." Champions are shared among brains to promote convergence of parameters in a similar locus. A process barrier is used as a synchronization step to allow $n=10$ training brains to receive $(n-1)=9$ champions. Once all champions are shared, the process barrier is set to "OFF" and the process continues. From the next generation set, the 3 dominant individuals are selected by tournament\cite{deb2002fast} and added to the champion set. Crossover with $cxp=1$ is applied to the extended champion set by matching even- with odd-positioned individuals, as in the preceding matching step. We process fitness values for the original and matched champions and a final selection of the best 8 individuals from the total set "next generation + original champions + matched champions" is used to upgrade the next generation set $M$. From $M$, in like manner, offspring are selected and the process continues for a new iteration.

Initially, the process explored several parameter values widely, and after several iterations, it gradually exposed a bifurcation of the inspection. Most of the parameters roughly followed an exploration path on each side of the default value. In order to decide which path leads to advancement of objectives, we compared objective values (Fig.\ref{fig:params_f}). The comparison helps to constrain exploration by reducing searching intervals toward better values and less computation time, speeding-up optimization. The new exploration lower $[0.01,0.32,0.01,0.01,0.1,]$ and upper $[0.10,0.65,0.13,0.22,3.0]$ bounds, achieved parameter stability after approximately the 20th iteration. Optimization stopped when slight changes in parameters produced almost no change in objective values, reaching $E=33$ iterations. Because optimization calculates fitness values twice for each iteration (3 times for the initial one, Fig.\ref{fig:process_moo}), the total number of iterations for each brain was $E^*=33\times2+1$. Every global tracking process takes 1 to 3 hours for the first several iterations; however, fiber density and length increase gradually while improving the parameters. Then every run becomes computationally expensive, sometimes lasting 10 or more hours.

\begin{figure}[h!]
\centering
\includegraphics[scale=0.22]{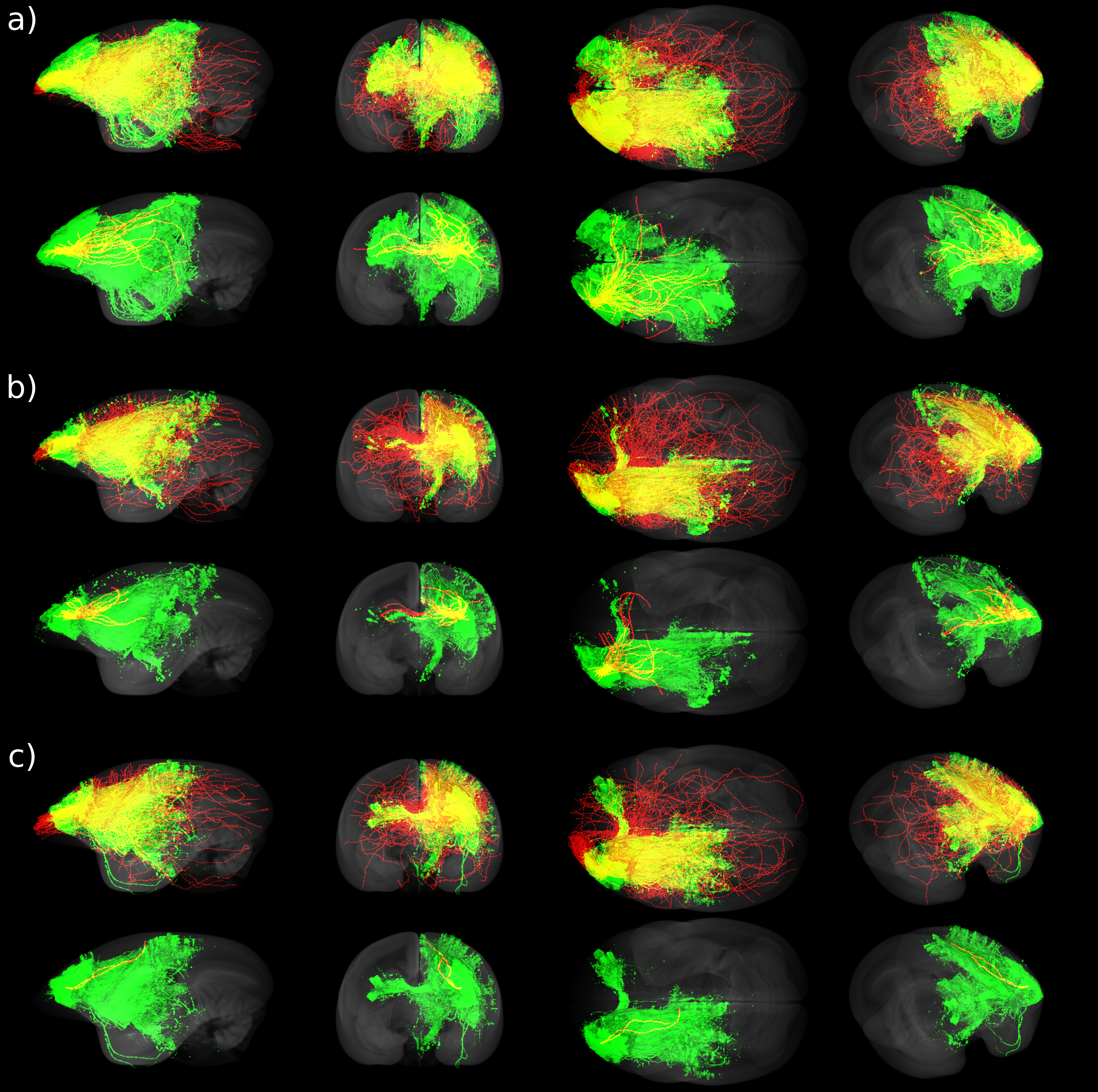} 
\caption{\textbf{Qualitative evaluation of optimization}. Unoccluded visualization of results of spatial relationships of dMRI-based, global tractography (red) and fluorescent tracer signals (green) of 3 injection sites a), b) and c) reconstructed by the data pipeline marmonet\cite{skibbe2019marmonet}.
Global tractography density maps are obtained from the intersection of full tractography results with injection regions. 
Their overlap (yellow) shows few common voxels for default parameters (lower rows). After optimization, global tracking shows improved results with enlarged overlap and longer fibers connecting sub-cortical and projection areas (upper rows).}

\label{fig:comparisons}
\end{figure}

\subsection*{Optimization results}

Results of fiber tracking with and without parameter optimization are visualized by overlapping dMRI-based fiber-density maps (red) with neural tracer data (green) (Fig.\ref{fig:comparisons}). Default settings (lower rows) generate sparse coverage, characterized by a few short fibers connected to the injection region. 
In contrast, tractography with optimized parameters (upper rows) presents expanded overlap with tracer signals, demonstrating higher sensitivity.  Longer fibers were connected not only to neighboring high-concentration neural tracer regions, but extended to cross-hemisphere areas and distant areas within the same hemisphere. 
The true/false positive ratio $f_2$ and the commissural passage $f_4$ allow control of the volatile growth of $FP$, while sensitivity and long-range connections are supported by the distance-weighted coverage $f_1$ and the projection coincidence $f_3$. 

\begin{figure}[h!]
  \begin{subfigure}{1.\textwidth}
    \centering
    \includegraphics[scale=0.75]{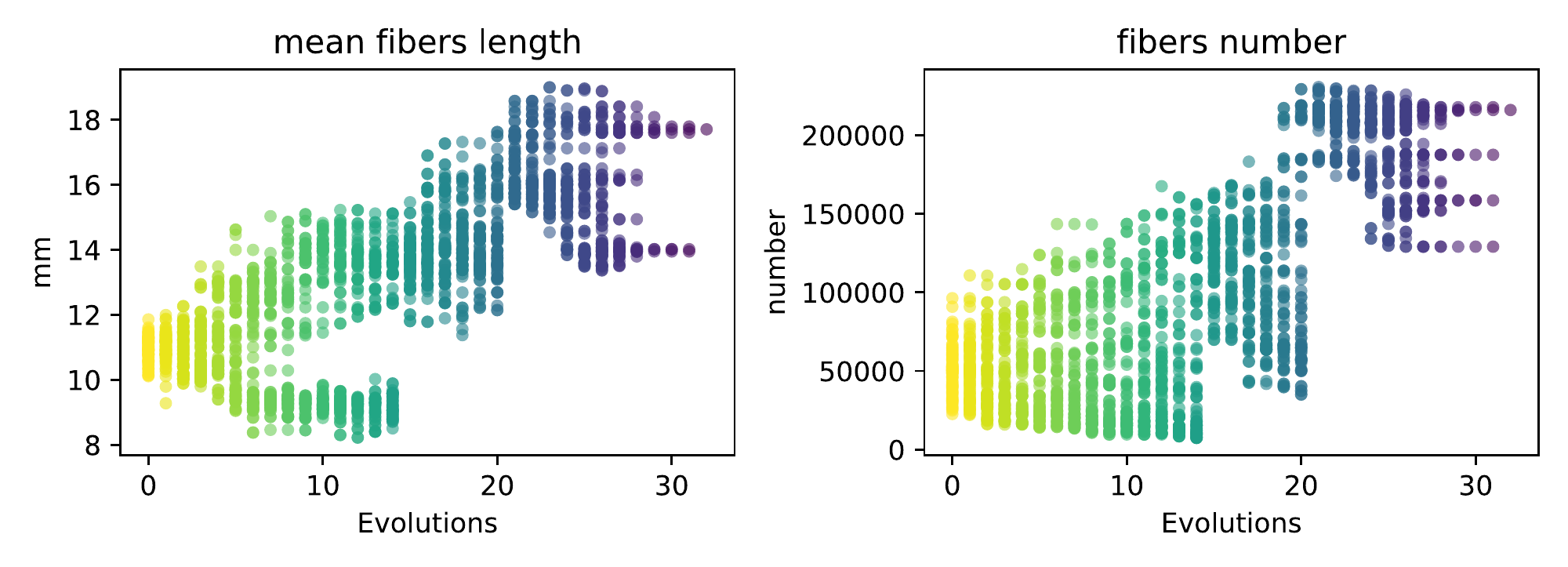}
    \caption{}
    \label{fig:mean_length_number}
  \end{subfigure}
  \begin{subfigure}{1.\textwidth}
    \centering
    \includegraphics[scale=0.3]{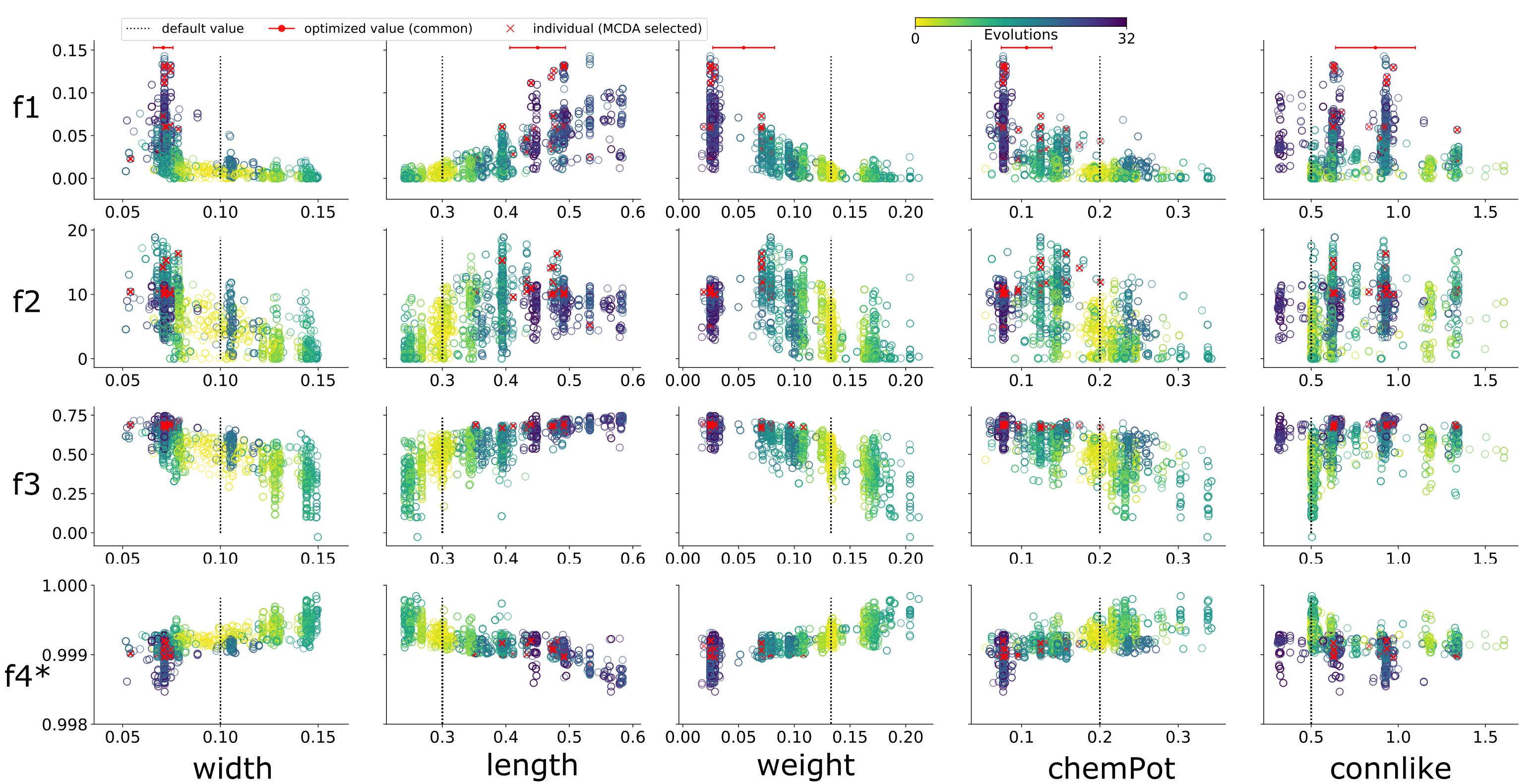}
    \caption{}
    \label{fig:params_f}
    \end{subfigure}
\caption{\textbf{Parameter exploration and convergence to generic values.} a) Evolution of the number and mean length of fibers through optimization. b) Given the default parameter values (black dotted line) and exploration ranges, optimization (color coded) widely scrutinizes potential values to maximize objective functions. Different exploration paths are observed, mainly because of different brains. $width$, $weight$ and $chemPot$ show robustness to multiple brains by converging to similar values. The generic setting is computed as the mean and standard deviation (red dots at the top row) of the best scoring parameters (red x markers).}
\label{fig:mean_length_number_params_f}
\end{figure}

We monitored the number and mean length of fibers estimated by the tractogram in the course of optimization (Fig.\ref{fig:mean_length_number}).
Both metrics increased from their default values of approximately $50,000$ fibers and $10$mm to optimized values of about $200,000$ fibers and $17$mm (see fiber length performance for a brain example at supplementary Fig.\ref{fig:length_hist}).
Higher fiber density helped to increase sensitivity in comparisons with tracer data, while longer fibers promote distant connections between source-target pairs. However, density increases must be constrained to avoid unrealistic results, controlled in our framework by $f_2$ and $f_4$. 

To verify the consistency and convergence of optimized parameters across subjects, we visualize evolution of the five parameters and four objectives for all ten training samples (Fig.\ref{fig:params_f}). 
Optimization started with parameters at their default values (Fig.\ref{fig:params_f}, dotted line) and  widely explored values within the defined search ranges. Over generations, parameters for all brains converged to similar loci while improving the objectives. Some parameters converged to almost the same value for all subjects, such as $width$, $weight$ and $chemPot$ (late iterations at Fig.\ref{fig:params_f}), whereas due to brain heterogeneity, $length$ and $connlike$ followed different paths to achieve the best results. This serves as an indicator of parameter robustness for generalization.

We chose standard parameters (the generic setting), using the mean and standard deviation of the best-scoring parameters by considering trade-offs between objectives (see MCDA below) and shown by red dots and bars in Fig.\ref{fig:params_f} and Table.\ref{tab:generic_params}.

\begin{figure}[h!]
\centering
\includegraphics[scale=0.292]{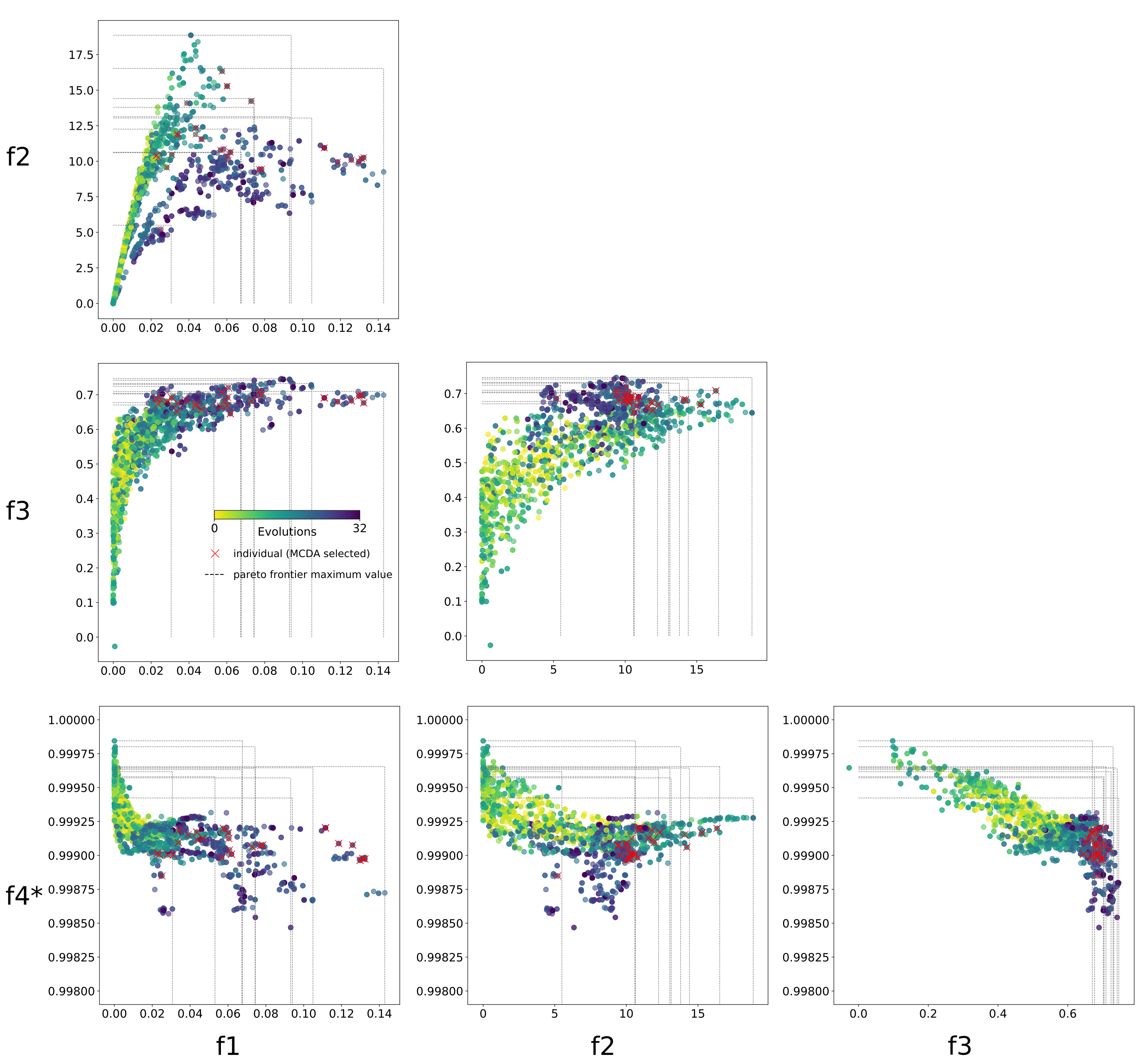} 
\caption{\textbf{Objective function optimization.} Pair-wise visualization of the optimization of the four proposed objective functions:
$f_1$: distance-weighted coverage, $f_2$: true/false positive rate, $f_3$: projection coincidence, and $f_4^*$: commissural passage.
NSGA-II drives objectives toward the pareto front in the upper-right direction. 
MCDA-based best objective trade-offs across brains are shown as red x markers. The standard setting is computed as their mean and standard deviation.}
\label{fig:f_f}
\end{figure}

To evaluate optimization of multiple objectives, we visualize the pair-wise evolution of the objectives (Fig.\ref{fig:f_f}). 

Different brains developed multiple pareto frontiers, which are most clearly seen in $f_1$ vs $f_2$ with dotted lines passing through the pareto's extremes (maximum value of $f$), which may be caused by subject individuality. 
However, systematic sharing of "champion" parameters enabled the algorithm to achieve optimal results in a similar locus among brains.

Competing goals $f_1$, $f_2$, and $f_3$ were "pushed" by the parallel optimization process from the lower-left (default parameters) to the upper-right region (optimized parameters) as seen in $f_1$ vs $f_2$, $f_1$ vs $f_3$ and $f_2$ vs $f_3$. 
$f_4^*$ maintained the proportion of valid fibers connecting hemispheres, a critical condition when the number of fibers increased (Fig.\ref{fig:mean_length_number}) and the tractography became denser. $f_1$ vs $f_4^*$, $f_2$ vs $f_4^*$, and $f_3$ vs $f_4^*$ indicate that $99\%$ of the crossing fibers passed through valid commissural voxels.

\subsection*{Choice of standard parameters by MCDA}

To assess trade-offs between objectives and to determine which combination performs best for each brain and for the whole set, we used Multiple Criteria Decision Analysis (MCDA) to select the standard set of parameters.

\begin{figure}[h]
\centering
\includegraphics[scale=0.7]{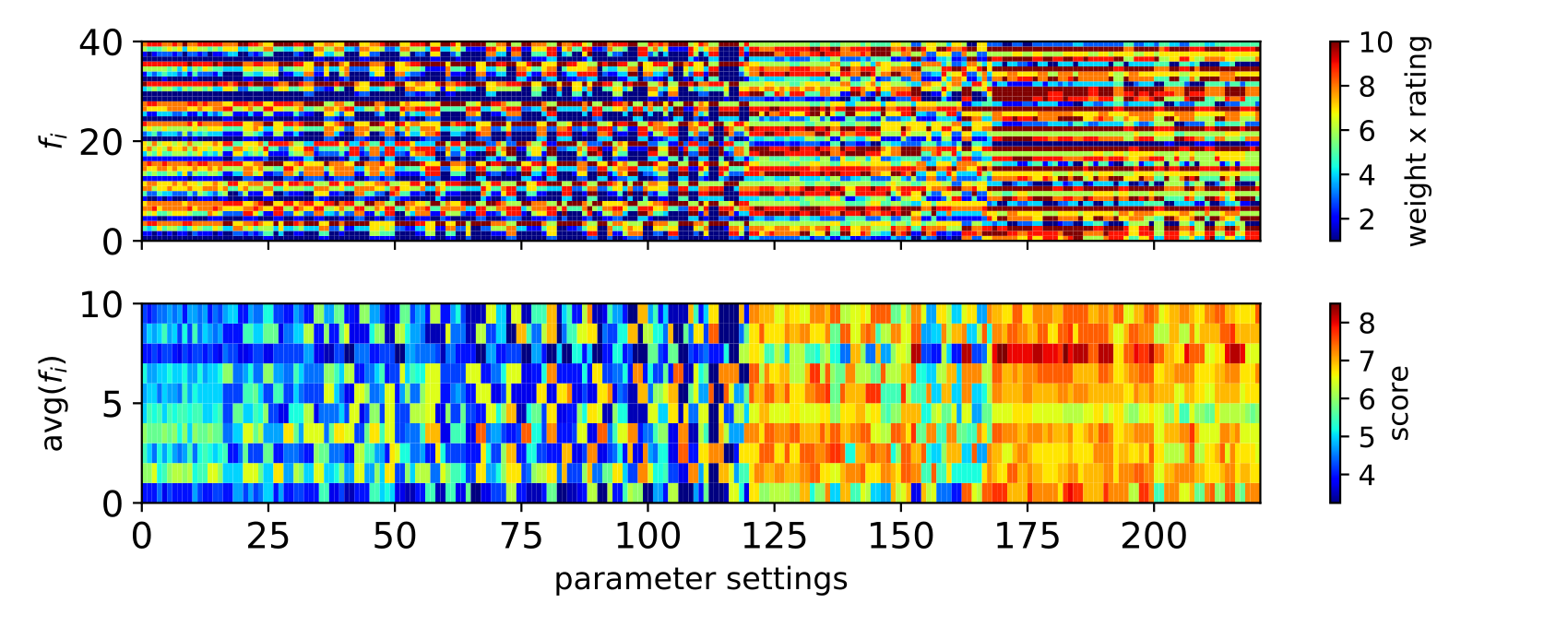} 
\caption{\textbf{Multiple Criteria Decision Analysis (MCDA).} Upper matrix rate parameter settings based on achieved objective values. The matrix below, averages the rates and calculates the final scores. Maximum scored settings per brain are selected as winners.}
\label{fig:mcda}
\end{figure}

For each brain, each objective $f$ is split in $10$ equal intervals in $[min(f),max(f)]$ and corresponding parameter settings are rated from $1$ (worst) to $10$ (best).
Ratings are arranged in a matrix of $40$ x $m$, where $40$ is the arrangement of the $4$ objectives for $10$ brains and $m$ is the number of parameter settings over the optimization (Fig.\ref{fig:mcda} upper matrix). 

Ratings are averaged across $f$'s with equal weighting for each brain, and the parameter set with the maximum score is selected as the winner(s) for the brain (Fig.\ref{fig:mcda}, lower matrix). 

Finally, the standard set of parameters is obtained using the mean and standard deviation of the winning parameters for the 10 brains. The result is shown in Table.\ref{tab:generic_params} with default parameters.

\begin{table}[h]
\caption*{\textbf{Parameters generalization}}
\normalsize
\centering
\begin{tabular}{|c|c|c|}
\hline
{Parameters ($\theta$)} & {Optimized value} & {Default} \\
{} & {(mean $\pm$ std)} & {value\cite{reisert2011global}} \\ \hline
$width$ ($\sigma$) & $0.07\pm0.005$ & $0.1$\\ 
$length$ ($l$) & $0.45\pm0.043$ & $0.3$\\
$weight$ ($w$) & $0.054\pm0.027$ & $0.133$ \\
$chemPot$ ($c$) & $0.106\pm0.032$ & $0.2$\\
$connlike$ ($L$) & $0.86\pm0.23$ & $0.5$\\
\hline
\end{tabular}
\caption{\label{tab:generic_params} Standard parameters for global tracking obtained by multi-objective optimization and MCDA over multiple marmoset brains.}
\end{table}

\begin{figure}[h]
\centering
\includegraphics[scale=0.7]{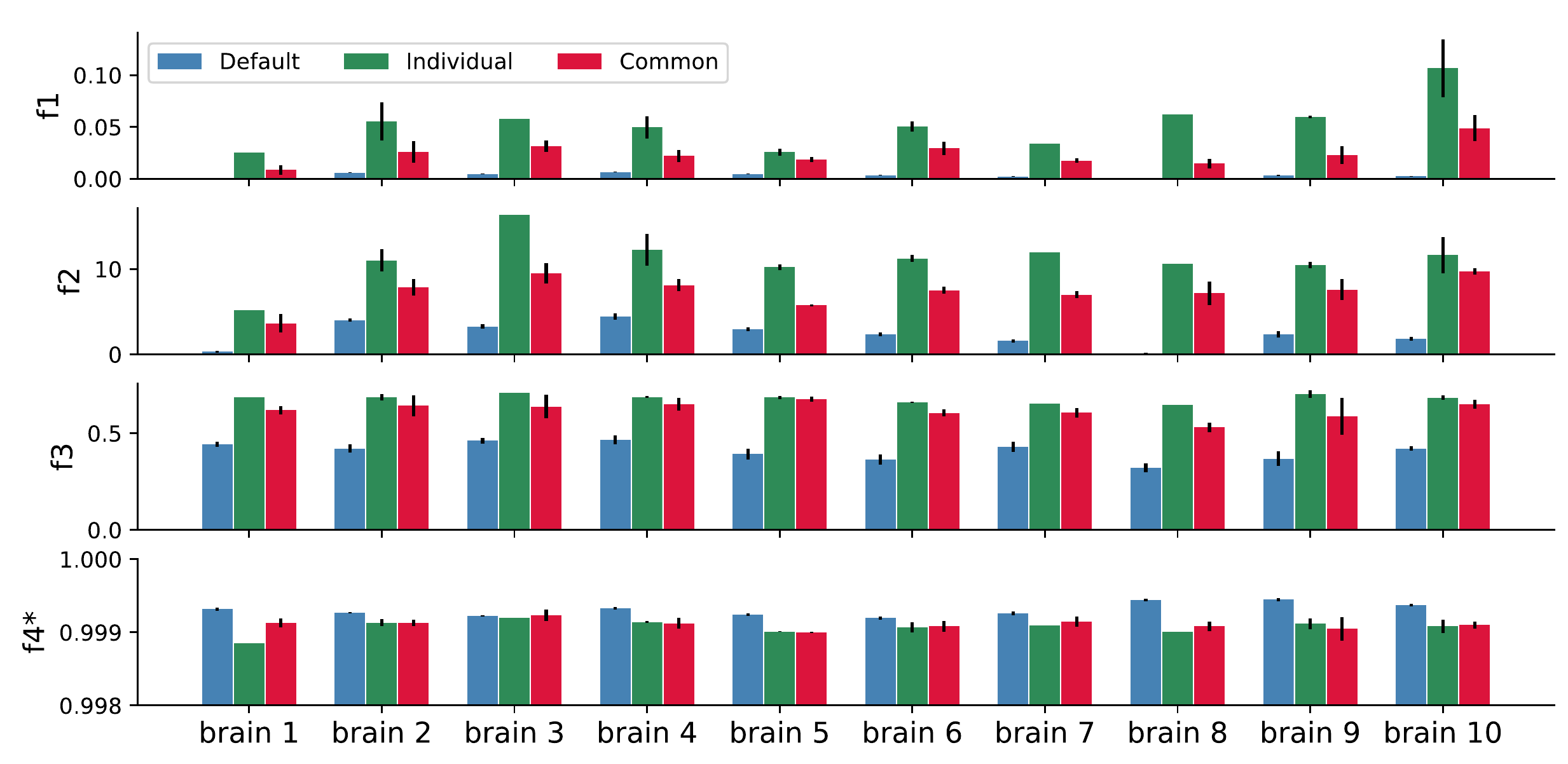}
\caption{\textbf{Training data set optimization results.} Objective function (average values for 5 runs) comparison of default, individual, and common (the latter two optimized) parameter settings for the training data set. Optimized parameters perform best at improving connections to projection areas $f_3$, while increasing coverage $f_1$ and true/false positive ratio $f_2$. Commissural passage $f_4$, expressed as the ratio of valid passage locations $f_4^*$, shows high accuracy.}
\label{fig:bars_known}
\end{figure}

\subsection*{Validation}

To validate the effectiveness of optimized parameters above, we compared training and test datasets in terms of the proposed objectives, for default and optimized parameters.
First, considering only the training set, we performed 5 global tractography runs for each default and optimized setting. In the latter case, each value is drawn from a normal distribution with its mean and standard deviation, as described in Table.\ref{tab:generic_params}. We averaged tractography run results for each brain (Fig.\ref{fig:bars_known}). We also show the performance of MCDA selected winners per brain, for comparison.

For individual winners and common standard parameters, $f_1$ obtained values of $0.067\pm0.036$ and $0.024\pm0.012$, $f_2$ values of $11.24\pm1.98$ and $7.38\pm1.88$, $f_3$ $0.68\pm0.016$ and $0.62\pm0.06$, and $f_4^*$ $0.99$ and $0.99$, respectively (Fig.\ref{fig:bars_known}). 
The standard parameters generalize well for improving cross-hemisphere projections ($f_3$) and commissural passage ($f_4^*$).
For $f_1$ and $f_2$, although the standard parameters achieved lower scores than the winners, they outperformed the default settings.
Compared to the results with default parameters, on average, $f_1$, $f_2$ and $f_3$ advanced from their low values ($0.003\pm0.002$, $2.3\pm1.4$ and $0.4\pm0.05$, respectively) to considerably better, optimized values (as shown above), reaching a better distance-weighted coverage $f_1$, while constraining false positives through $f_2$.
$f_4^*$ showed similar results for the three sets of parameters.
 
For the default case, coverage is low, and few fibers were generated, which leads to a high value of $f_4^*$. However, when $f_1$ increased by optimization, many more fibers were generated. A high value of $f_4^*$ indicates a similar level of accuracy at the commissural passage.

\begin{figure}[h!]
\centering
\includegraphics[scale=0.57]{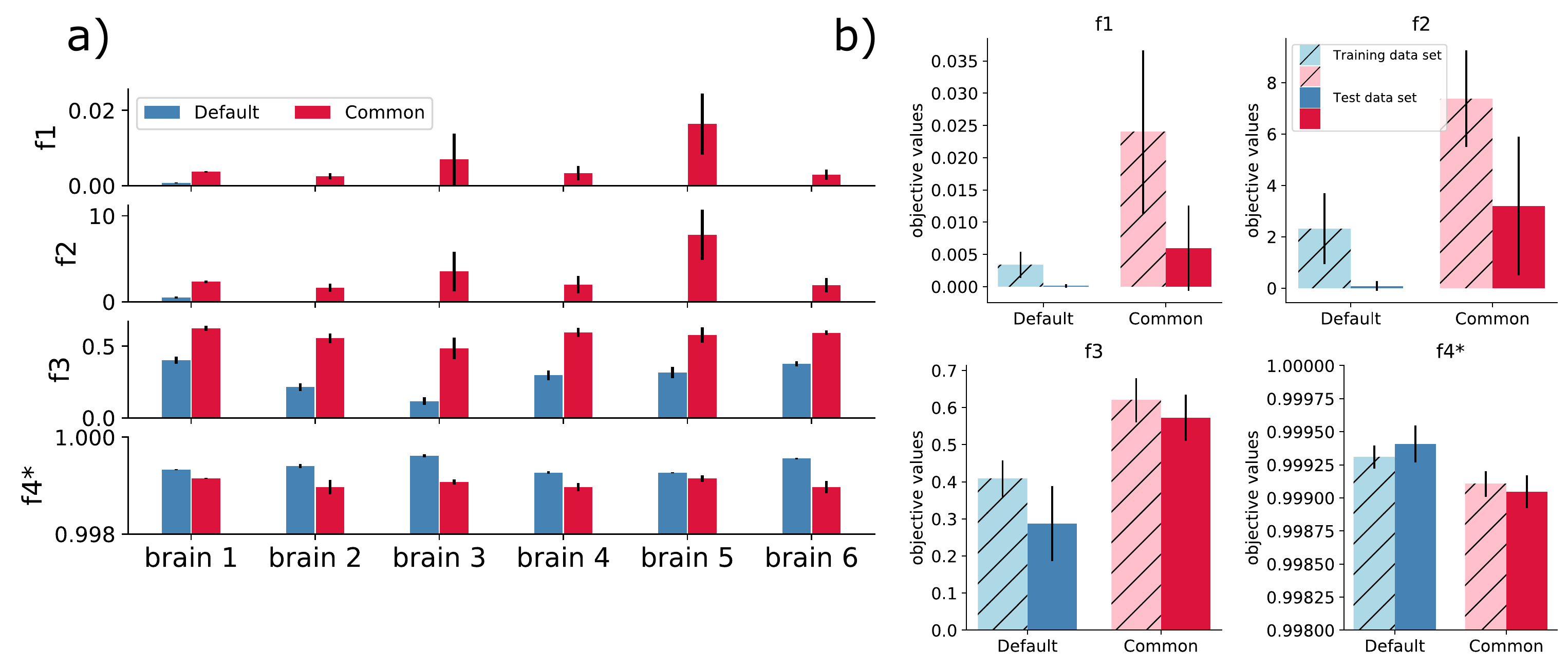}
\caption{\textbf{Performance on test data and comparison with training results.} a) Objective function comparison (average values for 5 runs) for 6 additional marmosets shows improvement of $f_1$, $f_2$ and $f_3$, and consistency of $f_4^*$. b) Performance comparison between training and test data sets for the default and optimized settings. $f_3$ is the most improved objective; however, improvement of $f_1$ and $f_2$ contributed to better results, as well as $f_4^*$ consistency for denser tractograms.}
\label{fig:bars_unknown_and_comparison}
\end{figure}

Generalization capability of the optimized parameters is also evaluated on 6 unseen marmoset brains (test set, Fig.\ref{fig:bars_unknown_and_comparison}a)). We ran tractography 5 times using the default parameters and the standard optimized parameters.
Results show improvement for $f_1$, $f_2$ and $f_3$, for all the brains. $f_1$ improved in average from $0.0001\pm0.0002$ to $0.006\pm0.006$, $f_2$ from $0.08\pm0.18$ to $3.2\pm2.7$ and $f_3$ from $0.28\pm0.1$ to $0.573\pm0.06$. As expected, $f_4^*$ showed similar results of about $0.99$.

Fig.\ref{fig:bars_unknown_and_comparison}b) summarizes the averaged performance for the training and test data sets, showing similar results. The objective $f_3$ shows better generalization performance.

Optimized parameters improved results in terms of the desired objectives for both cases, validating the proposed standard parameter settings. The improvements are clearly recognized in supplementary Fig.\ref{fig:tracer_individual_common} for a brain sample, which visualizes in high-resolution the ground-truth neuronal tracer signal (green) 3D reconstruction, and the global tracking fibers (red) in contact with the injection region, as density maps. Optimization improves fiber-density map matching with the neuronal tracer. Standard parameters perform similarly with decreased density results.

$f_1$ is affected by thousands of neural tracer voxels averaging the coincidences with voxels covered by global tracking fibers, and the mapping of fibers to a high-resolution space (standard brain). We evaluated the strength-weighted coverage $f_1^*=\frac{\sum_i^{N_{TP}} w_i}{\sum_i^{N_P} w_i}$ of axonal tracts at the voxel-level for the training set (see supplementary Fig.\ref{fig:axon_cov}) over the parameter settings of the optimization. The coverage improved in average from 0.9\% (default) to 15\% (MCDA selected winners).

\subsection*{Region-level connectomes}

Finally, we evaluated the region-level connectome matrix estimated using dMRI-based tractography in reference to the neural tracer-based connectome matrix. We took 20 injection points in the left prefrontal cortex to the rest of the brain, organized in a set of 500 ROIs (regions of interest).
This is the first version of a neural tracer-based connectome computed by  marmonet\cite{skibbe2019marmonet} in the Brain/MINDS program. 

Tractography-based matrices were mapped to the $20\times500$ structure. Both matrix components were log-transformed and normalized. 

\begin{figure}[h!]
\centering
\includegraphics[scale=0.45]{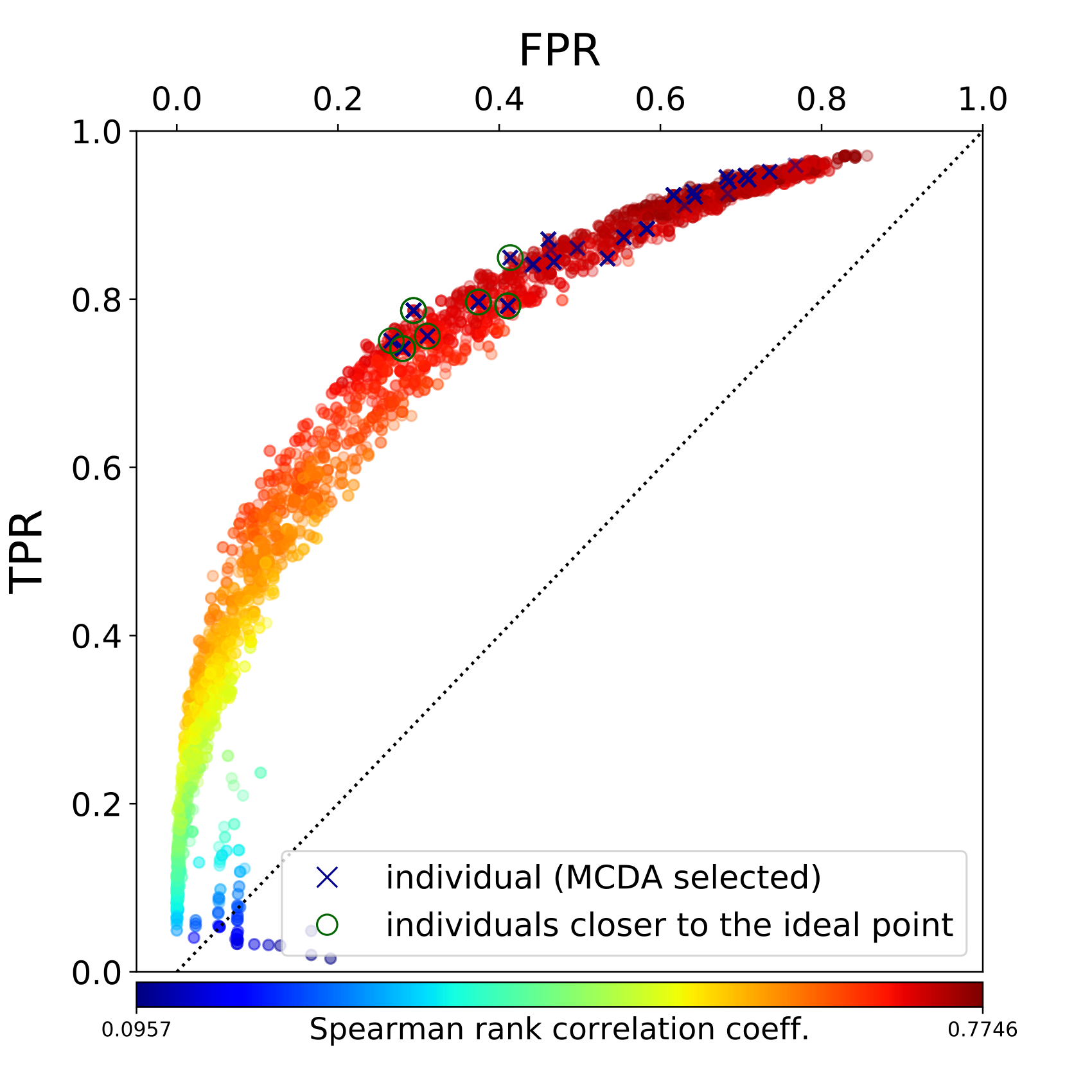}
\caption{\textbf{ROC space and correlation for dMRI-based matrices vs ground-truth connectome}. Spearman rank correlation coefficient space (color coded) between neural tracer and tractography-based matrices, mapped onto $TPR$-$FPR$ (brain-region level) curve over 500 ROIs for the entire optimization process. Optimized tractography results, as dark blue x's markers closer to the ideal coordinate (green circles) show high correlation.}
\label{fig:roc}
\end{figure}

Fig.\ref{fig:roc} visualizes the TPR-FPR space with Spearman rank correlation coefficients of the estimated and reference matrices in color.
Because dMRI-based tractography finds both incoming and outgoing fibers to and from a ROI, compared to an anterograde tracer-based connectome of only outgoing fibers, some "false" positives are reasonable.
Results of optimized parameters (blue x) substantially overlap settings close to the ideal point $(0.0,1.0)$ (green circles).
Optimized parameters achieved $FPR=0.33$, $TPR=0.78$, distance to the ideal point $d=0.163$, and correlation coefficient $r=0.724$ on average.

\begin{figure}[h!]
\centering
\includegraphics[scale=0.65]{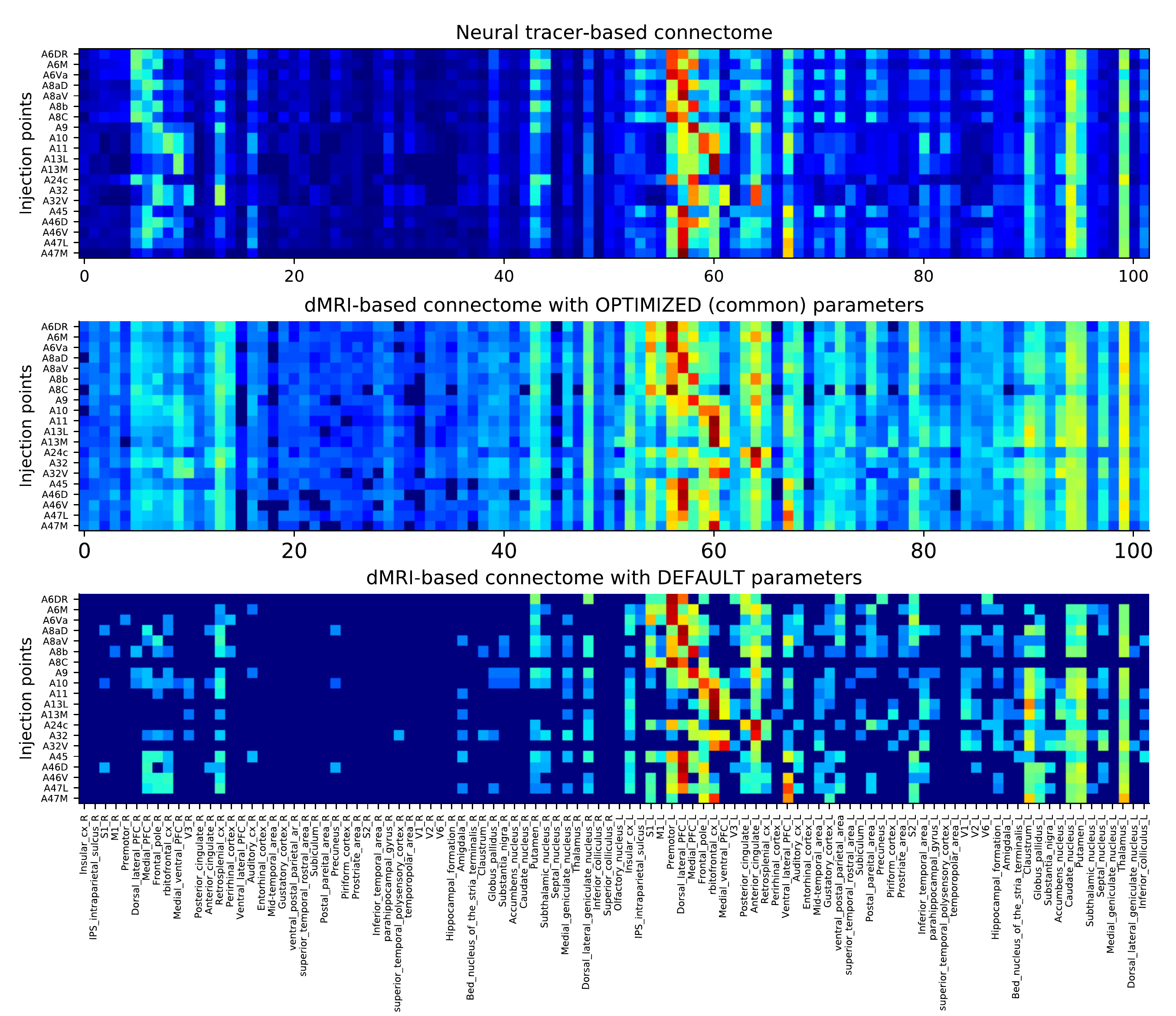}
\caption{\textbf{Brain-region level connectome comparison.} Preliminary neural tracer-based matrix (top) from marmonet\cite{skibbe2019marmonet} displaying relative connection strengths from 20 injection regions in the marmoset pre-frontal cortex to the rest of the brain, organized in 104 parcellations. For the sake of comparison, optimized (center) and default (bottom) dMRI-based connectomes for a brain-subject are mapped to the same structure.}
\label{fig:matrices_20x104}
\end{figure}

\begin{figure}[h!]
\centering
\includegraphics[scale=0.45]{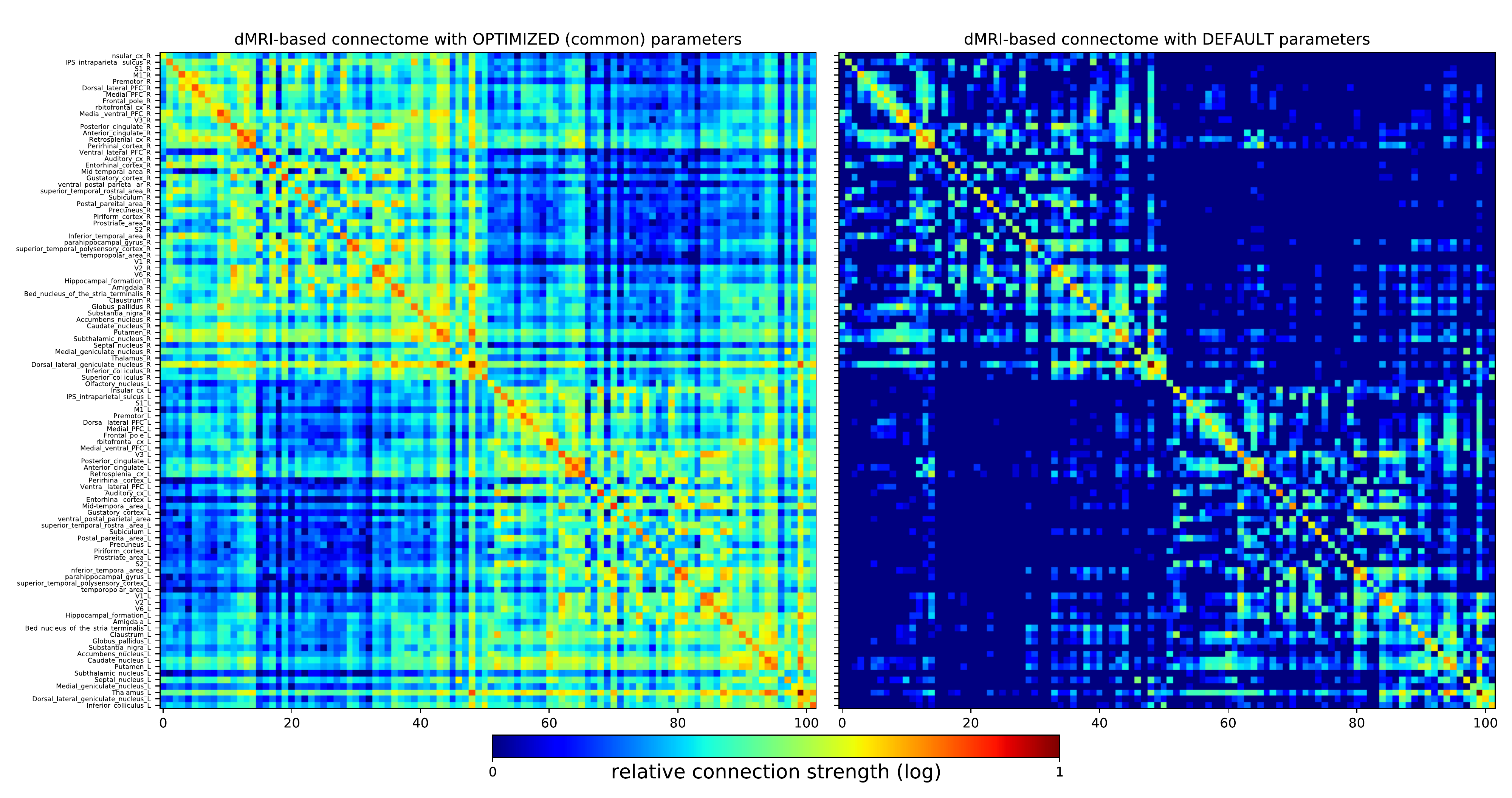}
\caption{\textbf{Whole-brain dMRI-based optimized connectome.} Square dMRI-based matrices comparison for one brain-subject example, using the optimized and default parameters.}
\label{fig:matrices_104x104}
\end{figure}

Improvements achieved with optimized parameters are easily recognized at the level of dMRI-based connection matrices. We organized the marmoset neural tracer-based matrix for 104 parcellations, and mapped global tractography results on the matrix of 20 injection points by 104 targets for the default and optimized settings (Fig.\ref{fig:matrices_20x104}). 
Compared to the sparse connections using default parameters (bottom matrix), tractography using optimized parameters (center matrix) revealed denser and longer connections, enhancing connectivity to projection areas in the right-hemisphere (left half of the matrices) from their origins in the left hemisphere.
Optimized dMRI tractography can complement the sparse structural network obtained from tracer injections (top matrix).
Optimization enhanced connectivity, not only from/to tracer-injected regions, but brain-wide (Fig.\ref{fig:matrices_104x104}), with richer connection estimations for the optimized case. 

\section*{Methods}

\subsection*{Fluorescent neural tracer data}
Neural tracer experiments and procedures were conducted with approval of the RIKEN CBS ethics committee. 3D tracer segmentation images (Fig.\ref{fig:process_moo}b, Fig.\ref{fig:comparisons}, Fig.\ref{fig:tracer_individual_common}a) are generated by marmonet \cite{skibbe2019marmonet}. Marmonet is the Brain/MINDS\cite{okano2016brain} AI-driven pipeline for automated segmentation of tracer signals. It incorporates state-of-the-art machine learning techniques based on artificial convolutional neural networks\cite{ronneberger2015u} and robust image registration. Raw images show the fluorescent signal of an anterograde tracer, a protein-based virus that tracks axons from injection region cells to their point of termination. Images are taken with a two-photon microscope, TissueCyte 1000 or TissueCyte 1100. Initially, they show several patterns, shapes, contrasts, and intensities. After marmonet pre-processing, image stitching, and segmentation, high-contrast results of the injection region and its center, corresponding cell bodies, and axon tracers are obtained. Segmentation results include voxel-intensity weighting from the raw tracer signal. We assume that axon densities and strengths of connections between injections and target regions are correlated with voxel intensity. All processed images are mapped from their $1.39 \times 1.34 \times 50 \mu m^3$ resolution to the Brain/MINDS reference image space\cite{woodward2018brain} of $100 \times 100 \times 200 \mu m^3$ resolution. Tracer injection regions and their centers as 3D reconstructions were used in our optimization as well.

\subsection*{Diffusion MRI}
Ex-vivo marmoset experiments and procedures were conducted with approval of the RIKEN CBS ethics committee. Marmosets were perfusion-fixed (Table.\ref{tab:MRI_brains}) and cranial brains were extracted. Brains were immersed in PFA reagent for 2-3 days, which was then replaced with PBS reagent. MRI imaging was performed on brains immersed in fluorinert liquid. A 9.4-Tesla small-animal MR scanner was used, controlled with Bruker paravision 6.0.1. The solenoid coil had an inner diameter of  $28 mm$. Diffusion imaging was accomplished using a spin-echo diffusion-weighted, echo-planar imaging sequence with repetition time $TR= 4000 ms$, echo time $TE = 21.8 ms$, and $b$-value $ = 5000 s/mm^2$. The acquisition matrix was $190 \times 190 \times 105$ over a $38 \times 38 \times 21 mm^3$ field-of-view (FOV), resulting in a native isotropic image resolution of $200 \mu m$. The diffusion sampling protocol included 128 unique diffusion directions and 2 non-diffusion-weighted (b0) measurements (The first b0 image was removed because it usually contains noise). Total acquisition time was 2 hours 40 minutes per sample.

\begin{table}[h]
\caption*{\textbf{Brains subjects used for optimization and validation.}}
\footnotesize
\centering
\begin{tabular}{|l|c|c|c|}
\hline
\multicolumn{4}{|c|}{\bf Training data set} \\
\hline
{Brain id} & {Gender} & {Fixed period (hours)} & {Age (years until the day of sacrifice)} \\ \hline

$R01\_0070\_CM1180F$ & $F$ & $80$ & $7$ \\
$R01\_0029\_CM696F$ & $F$ & $48$ & $6$ \\
$R01\_0072\_CM1176F$ & $F$ & $45$ & $11$ \\
$R01\_0030\_CM690F$ & $F$ & $48$ & $6$ \\
$R01\_0078\_CM1347F$ & $F$ & $95$ & $9$ \\
$R01\_0054\_CM1060F$ & $F$ & $60$ & $3$ \\
$R01\_0071\_CM1178F$ & $F$ & $143$ & $8$ \\
$R01\_0034\_CM521F$ & $F$ & $48$ & $3$ \\
$R01\_0039\_CM703F$ & $F$	& $48$ & $6$ \\
$R01\_0033\_CM694F$ & $F$ & $48$ & $6$ \\
\hline
\multicolumn{4}{|c|}{\bf Test data set} \\
\hline
{Brain id} & {Gender} & {Fixed period (hours)} & {Age (years until the day of sacrifice)} \\ \hline
$R01\_0026\_CM692F$ & $F$ & $48$ & $6$ \\
$R01\_0043\_CM628F$ & $F$ & $52$ & $4$ \\
$R01\_0040\_CM710M$ & $M$ & $72$ & $6$ \\ 
$R01\_0053\_CM1061F$ & $F$ & $58$ & $8$ \\
$R01\_0048\_CM1011F$ & $F$ & $60$ & $3$ \\
$R01\_0046\_CM1023M$ & $M$ & $60$ & $3$  \\
\hline
\end{tabular}
\caption{\label{tab:MRI_brains} Characteristics of marmoset brains used in this study. The same brains were handled for tracer injections and dMRI imaging.}
\end{table}

\subsection*{Pre-processing}
dMRI data, bvec and bval files, and individual whole-brain masks were acquired from the Brain/MINDS dMRI-pipeline. dMRI was de-noised using MRtrix3\cite{tournier2019mrtrix3} in 3 steps. First we applied $dwidenoise$, which exploits data redundancy in the PCA domain using random matrix theory\cite{veraart2016denoising,veraart2016diffusion}; secondly $mrdegibbs$ removed Gibbs ringing artefacts by local subvoxel-shifts\cite{kellner2016gibbs}. Finally, a mask filter was applied to the whole-brain mask, eroding 2 voxels to remove noise at the boundaries and to constrain abnormal fiber growth during global tracking. Injection region masks\cite{skibbe2019marmonet} were dilated 2 voxels to improve detection of fibers contacting them, as support against potential bias in the registration and injection region detection. For registration tasks we used b0 images and advanced normalization tools ANTs\cite{avants2011reproducible}.

\subsection*{Connectome data}
Comparisons at the brain-region level used a preliminary neural tracer-based structural connectivity matrix from marmonet. It integrates tracer data from several marmoset brains into a connectome of 20 injection regions and 500 projection targets given by the Brain/MINDS atlas. Additionally, the neural tracer-based connectome is mapped to a lower level of granularity, a matrix of 20 injection regions and 104 parcellations. 

dMRI-based matrices were created for each global tracking result in standard brain space. The connectome was built by assigning each streamline to all regions it intersects. Neural tracer and dMRI-based matrices were log-transformed and normalized before comparison.

The neural tracer connectome is unidirectional, from the left hemisphere (injection points) to the right hemisphere (targets); however, in order to compare with tractography, we assumed that regions are connected independently of tracer directionality. The Brain/MINDS marmoset connectivity map is an on-going effort. At the time of this report, the first results correspond to injections in the left pre-frontal cortex of the marmoset.

\subsection*{Density maps}
Evolutionary optimization requires neural tracer data and a comparison of fiber-density maps in standard brain space (Fig.\ref{fig:criteria}). A fiber-density map is built for each individual (a particular parameter setting). First, duplicated global tracking results are transferred from dMRI space to standard brain space by normalization mapping ($tcknormalise$ from MRtrix3\cite{tournier2019mrtrix3}). In that latter space, tractograms are intersected with the corresponding tracer injection region using $tckedit$\cite{tournier2019mrtrix3}. The sub-set of fibers in contact with the injection region, as well as the complete set of fibers, is converted to density maps by $tckmap$\cite{tournier2019mrtrix3}, and averaged over the duplicated tractography runs. The sub-set of the fiber density map is compared in 3D space at voxel-level with the neural tracer signal reconstruction, so $f_1$ and $f_2$ can be obtained. Similarly, $f_4^*$ is measured by the intersection of the transverse mask with the full set of fiber density maps.

\subsection*{Voxel weighting}
Each voxel of $f_1$ is weighted with 2 factors obtained from neural tracer data, the distance $d_i$ and intensity $w_i$ (Fig.\ref{fig:criteria}). The center of the injection region is composed of a few voxels. A refinement to a unique voxel is performed by summing all x, y and z-coordinates and dividing each sum by the corresponding number of voxels, giving a unique 3D position. The updated center is used to calculate the distances $d_i$ from all $TP$ voxels, to the injection center. Distances $d_i$ are normalized by the maximum observed distance. Raw neural tracer 3D images provide voxel intensities $w_i$, which are associated with connection strengths from the injection site to the projection areas\cite{skibbe2019marmonet}. Similarly, $w_i$ is normalized by the maximum observed intensity.

\subsection*{Global tractography and parameter selection}
dMRI-based tractography was performed using a global tracking algorithm\cite{reisert2011global}. This method provides the whole-brain connectivity configuration that optimally fits the acquired data \cite{reisert2011global,mangin2013toward,christiaens2015global}. The optimization applied is such that each particle (also called a segment) tries to mimic the source data, promoting its closeness to the measurement in anisotropic areas (e.g. the white matter), and infers information in ambiguous isotropic areas (e.g. gray matter) using neighboring anisotropic areas. We selected this algorithm due to its documented reliability in terms of position, tangent directions, and curvature of reconstructed fibers with a phantom dataset at the DMFC-fiberCup at MICCAI'2009. However, it requires optimization for specific anatomy or species. Nevertheless, the proposed multi-objective optimization method can be applied to any tractography algorithm.\\
Global tracking does not use pre-defined seed(s), requiring no human intervention. Fibers are built with small line segments that form chains during tractographic optimization, and their number and orientation are adjusted to match data obtained from high angular resolution diffusion imaging (HARDI). From the set of segments and their connections, a predicted MR-signal is computed. Connection behavior between segments is controlled by internal energy, from two parameters selected as relevant to our optimization. $length$ $l$ is the fiber segment length and $connlike$ $L$ is the likeliness that two segments link together (also known as connection potential). An external energy measures the difference between the current and predicted diffusion-weighted HARDI signals. From the external energy we designated the $weight$ $w$ contribution and the $width$ $\sigma$ of the prototype-signal of each segment as important parameters. In addition, two more parameters were considered, the $chemPot2$ $c$ (cost of adding a particle) and $chemPot1$ (similar to chemPot2, also known as the particle potential, which regulates the number and distribution of particles).\\
To test the significance of the selected parameters, we pre-evaluated them by running global tracking on 3 brains and by assessing the fiber number and length variability caused by a single parameter change, while keeping others fixed at their default values (supplementary Fig.\ref{fig:gt_params_selection}). $Weight$, $width$, $length$ and $connlike$ parameters produced changes in fiber density and length. However, for $chemPot2$ and $chemPot1$ cases, changes of the parameter value produced almost no effect on fibers density and length, and was practically unnoticeable in the latter case. Therefore, we selected the first 4 parameters and $chemPot2$ (renamed as $chemPot$) for optimization.

\subsection*{Code implementation}

The method reported here was implemented on a cluster HPC computer. In order to share champion settings over each iteration, we ran separate jobs for each brain, and synchronized them. There are 2 types of jobs, light (synchronized) and heavy. Light jobs keep running the evolutionary processes for the brains (1 job per brain). They were tested on a single core with low memory; however, they are active during the whole optimization process, which took around 4$\sim $5 weeks.

For each iteration, a light job creates and dispatches 8 heavy jobs (1 job per individual parameter setting). A heavy job uses more than 1 core and requires higher memory. Heavy jobs read data sources (masks, neural tracer reconstructions, dMRI, atlas, injection regions), perform global tracking $n$ times, calculate objective functions, and record results (jobs information, parameters, tractograms, density maps, champions, connection matrices, objectives value) in a folder-organized structure.

By this method, optimization process parallelization is implemented at the level of individual brains and global tracking runs.

\section*{Summary}

We optimized and validated parameters of the global fiber-tracking algorithm \cite{reisert2011global} by exploiting fluorescent tracer and dMRI data from the same marmoset brains in the Brain/MINDS Project \cite{okano2016brain}.

To address the competing goals of sensitivity and specificity for multiple brains, we took a parallel, multi-objective optimization framework with four objective functions (Fig.\ref{fig:criteria}); two voxel-level objectives ($f_1$: distance-weighted coverage, $f_2$: true/false positive ratio), a region-level objective ($f_3$: projection coincidence), and an anatomical constraint ($f_4$: commissural passage). 

Optimization was based on a NSGA-II evolutionary approach and implemented champion parameter sharing across brains to promote parameter generalization while maximizing the objectives (Fig.\ref{fig:process_moo}).

During the optimization process, while constraining impossible fibers at the commissural passage and controlling the growth of false positives, our framework improved dMRI-based fiber tracking performance with respect to the default values: average fiber length from 10mm to 17mm (Fig.\ref{fig:comparisons}, \ref{fig:mean_length_number_params_f} and \ref{fig:length_hist}), voxel-wise coverage of axonal tracts from 0.9\% to 15\% (Fig.\ref{fig:axon_cov}), and correlation of target areas from 40\% to 68\% (Fig.\ref{fig:bars_known}).

From the multiple pareto-optimal solution for multiple brains, we used a MCDA method to select a standard set of parameters (Table.\ref{tab:generic_params}).

Using brain samples that were not used for optimization as part of the test set, we verified that the standard parameters substantially improve fiber tracking performance compared to the default parameters (Fig.\ref{fig:bars_unknown_and_comparison}, \ref{fig:roc}, \ref{fig:matrices_20x104} and \ref{fig:matrices_104x104}).

These results also raise concerns about dMRI-based connectome studies that lack optimization or validation of fiber-tracking algorithms.

Recently, Zhang et al.\cite{zhang2018optimization} proposed optimization of dMRI-based fiber tracking using the region-level coincidence with neural tracer data in the CoCoMac database\cite{bakker2012cocomac} and the matching of fibers orientations  with myelin stained data from one macaque brain\cite{mikula2007internet}. They took the average of Youden’s index (sum of sensitivity and specificity)\cite{youden1950index} for the connected regions and the coincidence index of fiber orientation as the criterion and performed a grid search in a two dimensional parameter space of a fiber-tracking algorithm\cite{yeh2010generalized}.
Our framework, using a multi-objective optimization algorithm, can flexibly incorporate multiple evaluation criteria and optimize a larger number of parameters. Another important feature of our work is that the comparison of the dMRI data and tracer data are are performed in parallel for multiple brains, which can take into account individual variability.

Originally, we started this effort by optimizing a single objective function, such as $C^2 = FPR_v^2+(1-\frac{\sum_i^{N_{TP}}d_i}{\sum_i^{N_P}d_i})^2$, where the second term is the normalized sum of distances from $TP$ voxels to the center of mass of the injection region, similar to $f_1$, but using only $d_i$ as a weighting factor. 
However, the results for the combined single objective function by the co-variance matrix-adaptation evolution strategy (CMA-ES)\cite{hansen2011cma, hansen2016cma} were unsatisfactory, with a huge density of fibers, dominance of false positives, and many fibers crossing hemispheres outside the commissures.

Better objective functions may exist, so we encourage a search for more efficient objectives within a multi-objective framework. For example, we performed a $TPR$-$FPR$ at brain-region level evaluation, which could be added as an objective function. Moreover, our multi-objective optimization framework can be applied to other fiber-tracking algorithms.



Data and tool sharing are crucial. That is why projects like Brain/MINDS Project\cite{okano2016brain}, among several others, are so important. Our implementation code is available to the scientific community for improving dMRI-based fiber-tracking accuracy and reliability.

\section*{Author contributions}
CEG, HS and KD wrote the paper. CEG implemented and ran the optimization process, and analyzed data. CEG and HS designed objective functions and performed the coding. HS performed tracer signal segmentation and 3D brain reconstructions for comparisons, provided the transverse mask and the tracer connectome. KN and HT contributed the dMRI pre-processing, individual brain masks and mapped dMRI to the reference space. CEG and JL designed the optimization process. JH and HO provided the dMRI. MR and HS contributed the global tracking algorithm and adapted it for command line execution. AWa and TY contributed the anterograde tracer-injected brain raw data. AWo and YY provided the standard average brain and the atlas. KD, SI, CEG, and HS designed the research.

\section*{Data availability}

 The optimization process code is publicly available on github (https://github.com/oist/gt\_moo/). It can be adapted to other fiber tracking algorithms and data sources.

Global tracking algorithm is available at https://www.uniklinik-freiburg.de/mr-en/research-groups/diffperf/fibertools.html .

Data sources (neural tracer, dMRI, standard brain, atlas, neural tracer connectome, masks) are not publicly available, however, they will be gradually disclosed as part of Brain/MINDS project in the near future (data portal site: https://www.brainminds.riken.jp).

Any other data presented in this study are available from the corresponding author upon reasonable request.

\section*{Acknowledgements}
This research was supported by the program for Brain Mapping by Integrated Neurotechnologies for Disease Studies (Brain/MINDS) from the Japan Agency for Medical Research and Development AMED (JP17dm0207030 to KD, JP20dm0207001 to HS, JP19dm0207088 to KN), the KAKENHI Grant 16H06563, the Post-K Exploratory Challenge 4 ”Understanding the neural mechanisms of thoughts and its applications to AI“ of MEXT (Ministry of Education, Culture, Sports, Science and Technology of Japan) to KD, and internal funding from the Okinawa Institute of Science and Technology Graduate University. We thank Steven D. Aird for editing the manuscript.

\bibliography{moo_ref}

\pagebreak

\beginsupplement

\subfile{suppmaterial.tex}

\end{document}

%% file: suppmaterial.tex
\section*{Supplementary materials}

\begin{figure}[h]
\centering
\includegraphics[scale=0.75]{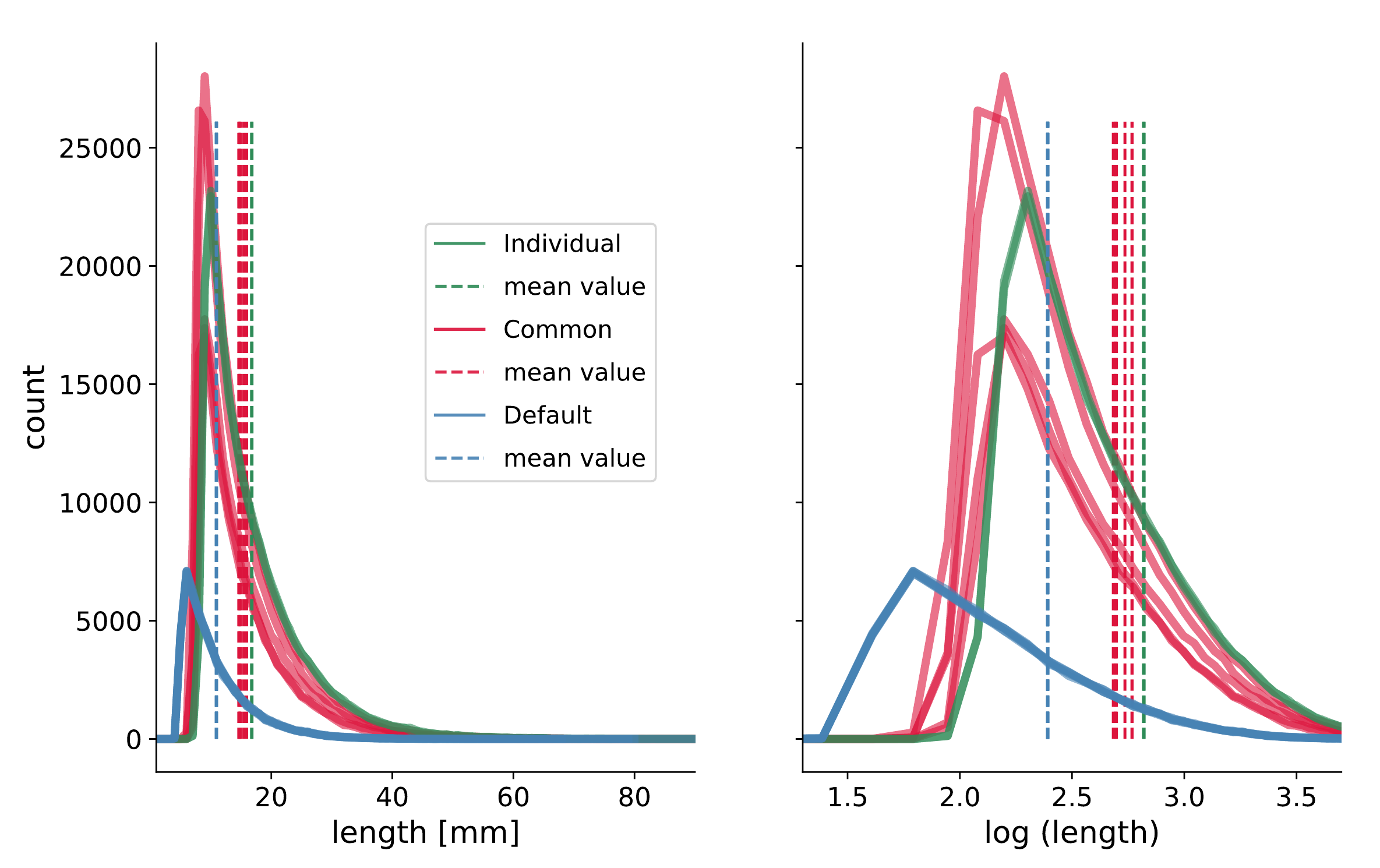} 
\caption{\textbf{Fiber length performance.} Fiber length histograms for a brain subject showing the default and optimized (individual and common) parameter results. 5 runs for the default and standard (common) parameters show the improvements, better displayed in the log-transformed histograms (right figure). Optimization increased considerable the number of fibers, and extended their lengths. The mean fiber length (dashed lines) improved from  around 10mm to 17mm. Common results are similar to the performance of the best setting (individual) found for the subject.}
\label{fig:length_hist}
\end{figure}

\begin{figure}[h!]
\centering

\includegraphics[scale=0.355]{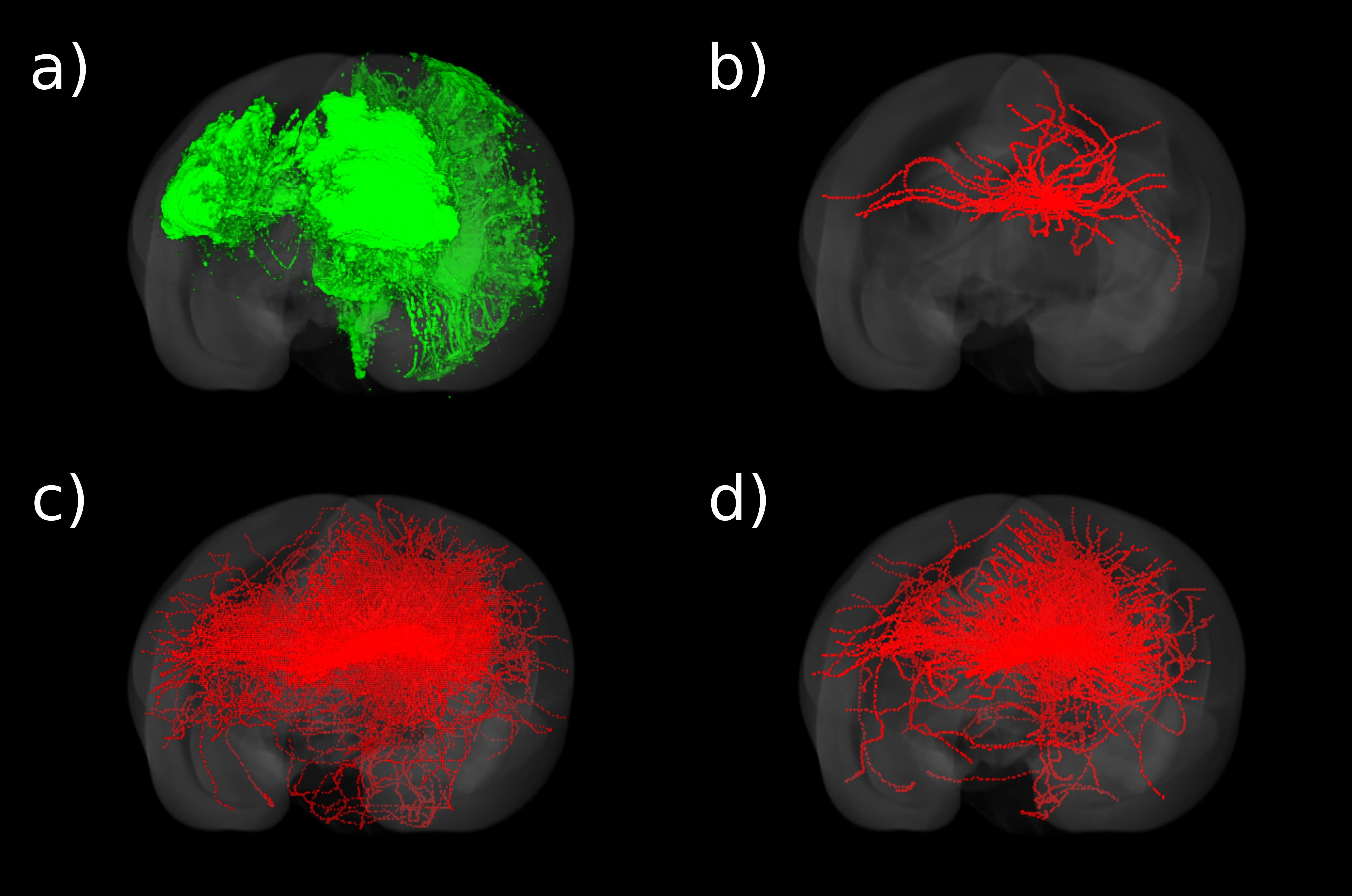} 
\caption{\textbf{Optimized and default dMRI fiber tracking results.} Visualization of a sample ground-truth neuronal tracer signal (green) 3D reconstruction a) and global tracking results (red) using different parameters settings. b) Fibers in contact with the injection region as a density map for the default parameters. c) Optimization improves fibers density map matching with the neuronal tracer for the individual setting. d) Standard parameters perform similarly to c), providing less dense results.}
\label{fig:tracer_individual_common}
\end{figure}

\begin{figure}[h]
\centering
\includegraphics[scale=0.75]{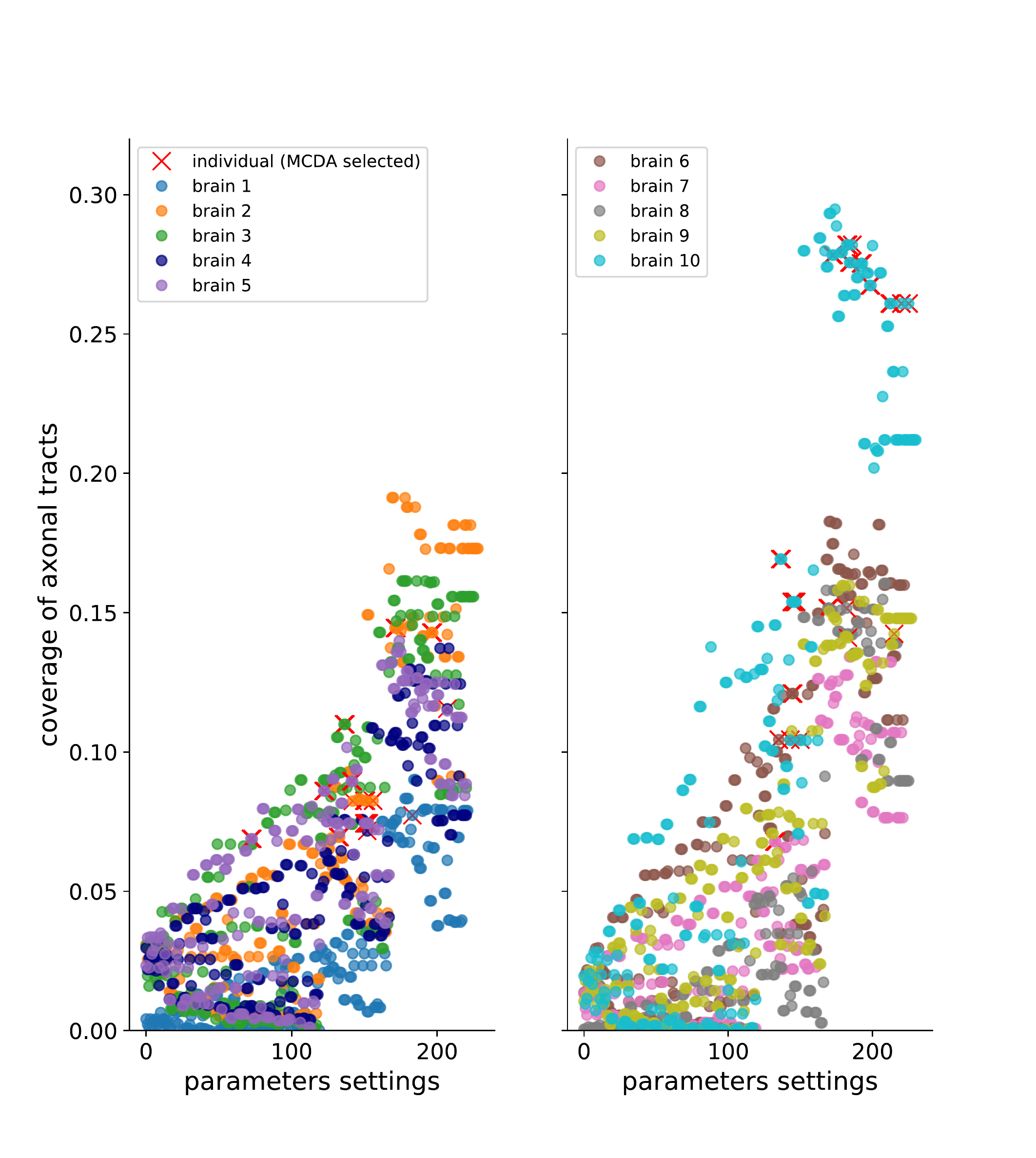} 
\caption{\textbf{Coverage of axonal tracts.} Performance of the strength-weighted coverage ($\sum_i^{N_{TP}} w_i/\sum_i^{N_P} w_i$) of axonal tracts (neural tracer) by global tracking fibers for the 10 brains of the training set. Coverage improved in average from $0.9\%$ (default settings) to $15\%$ (winners settings, red x markers). A subject with less tracer volume obtained about $30\%$ of coverage (brain 10). The values of the coverage are affected by the high number of the ground-truth positive voxels and the mapping of fibers to high-resolution space.}
\label{fig:axon_cov}
\end{figure}

\begin{figure}[h]
\centering
\includegraphics[scale=0.6]{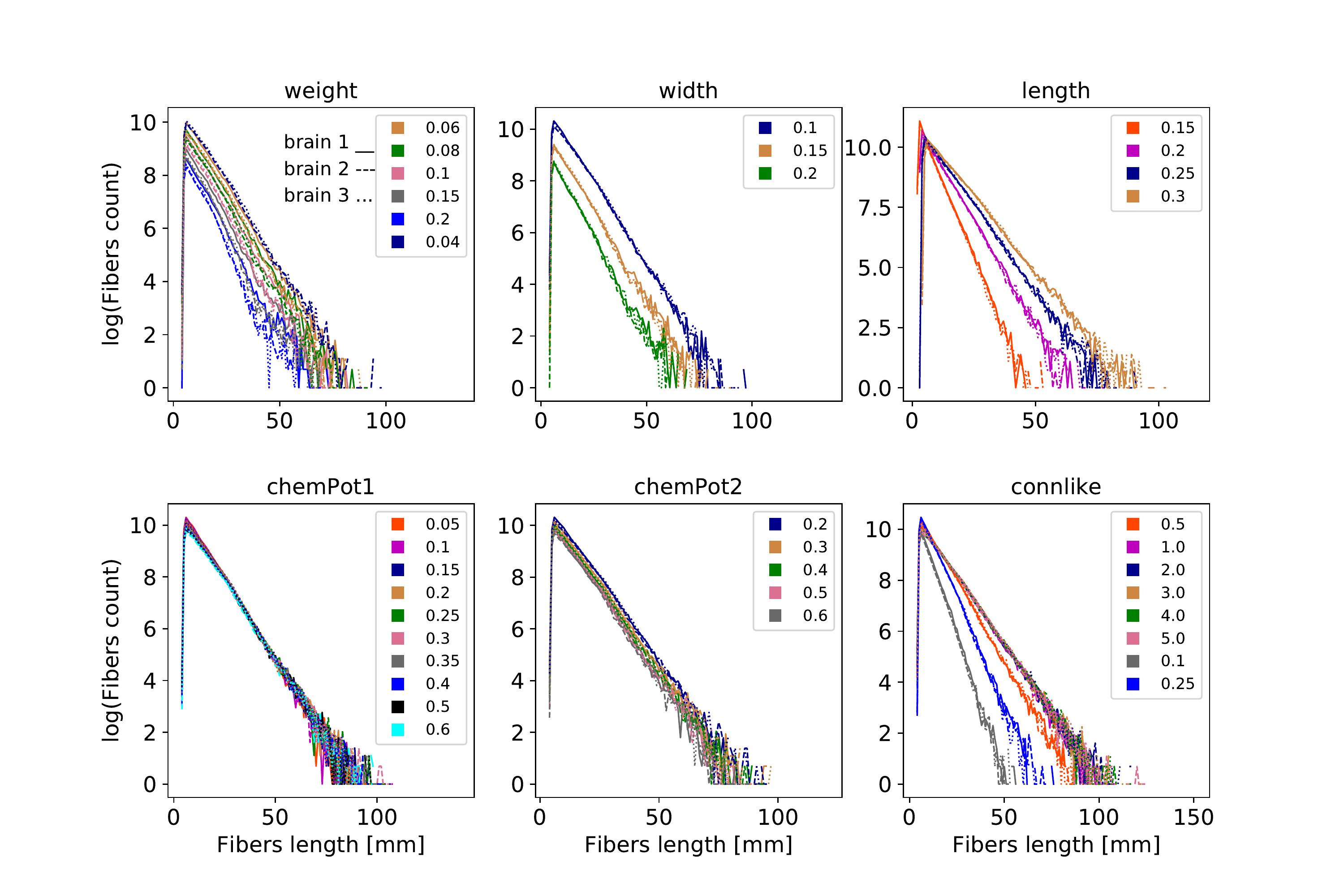}
\caption{\textbf{Parameters selection by relevance.} Fibers length histograms for 3 marmoset brains global tracking results. Varying one parameter while maintaining fixed the others provides a clue about parameter's relevance. Bottom-left sub-plot shows almost no change for different values of $chemPot1$.}
\label{fig:gt_params_selection}
\end{figure}